\documentclass[10pt]{article}
\usepackage[a4paper,margin=1.6cm,columnsep=0.7cm]{geometry}
\usepackage[T1]{fontenc}
\usepackage{lmodern}
\usepackage{authblk}
\usepackage{graphicx}
\usepackage{booktabs}
\usepackage{array}
\usepackage{longtable}
\usepackage{pdflscape}
\usepackage{textcomp}
\usepackage{caption}
\usepackage{tabularx}
\usepackage{xcolor}
\usepackage[numbers,sort&compress]{natbib}
\usepackage[hidelinks]{hyperref}
\setcitestyle{numbers,square}
\graphicspath{{./figures/}}

\renewcommand{\arraystretch}{1.12}
\newcolumntype{L}[1]{>{\raggedright\arraybackslash}p{#1}}
\renewenvironment{abstract}
  {\par\noindent\textbf{ABSTRACT}\par\vspace{0.2em}\noindent}
  {\par}

\begin{document}

\title{THz Spectroscopy of Urine Vapors from Patients with Prostate Cancer and Benign Prostatic Hyperplasia: A Critical Review and Statistical Reanalysis}

\author[1,2]{V. A. Atduev}
\author[2,3]{A. V. Maslennikova}
\author[3,4]{V. L. Vaks}
\author[2]{A. P. Bavrina}
\author[3,4]{E. G. Domracheva}
\author[3,4]{M. B. Chernyaeva}
\author[3,4]{V. A. Anfertev}
\author[1]{K. A. Atduev}
\author[5]{Y. Tawalbeh}
\author[5]{M. F. Pereira}

\affil[1]{Privolzhsky District Medical Center of the Federal Medical and Biological Agency of Russia, 2 Nizhnevolzhskaya Embankment, Nizhny Novgorod 603001, Russia}
\affil[2]{Privolzhsky Research Medical University, 10/1 Minin and Pozharsky Square, Nizhny Novgorod 603005, Russia}
\affil[3]{Lobachevsky State University, 23 Gagarina Avenue, Nizhny Novgorod 603022, Russia}
\affil[4]{Institute for Physics of Microstructures of the Russian Academy of Sciences, GSP-105, Nizhny Novgorod 603087, Russia}
\affil[5]{Department of Physics, Khalifa University of Science and Technology, Abu Dhabi 127788, United Arab Emirates}

\date{}
\maketitle

\begingroup
\fontsize{9}{10.5}\selectfont
\begin{abstract}
The cancer specificity of serum prostate-specific antigen (PSA) motivates complementary, non-invasive approaches to prostate disease assessment. Building on reported high-resolution terahertz (THz) analysis of urine-derived products, this review integrates spectroscopic evidence with a statistical reanalysis of clinical data from 24 patients with prostate cancer (PC) and 14 with histologically confirmed benign prostatic hyperplasia (BPH). Nonparametric statistical comparison shows higher PSA levels in PC, while overlapping patient distributions and receiver-operating-characteristic analysis with bootstrap uncertainty estimation demonstrate incomplete discrimination from BPH. Fast frequency-sweep spectroscopy characterizes urine-derived volatile and thermal-decomposition products, distinguishing compounds shared between groups from candidate PC-associated assignments, including glycolaldehyde, butyronitrile, pentanenitrile and methyl isocyanate. Glycolaldehyde has been detected by urinary carbonyl profiling, while benzaldehyde and phenol have been reported in urinary volatile studies. Related aldehydes identified by gas chromatography-mass spectrometry and glycine-associated metabolites reported in proton nuclear magnetic resonance studies provide complementary chemical and metabolic context. These correspondences support plausibility of THz findings without establishing PC specificity. Together, the dataset, statistical assessment and comparative literature synthesis support THz urine spectroscopy as a platform for molecular discovery and candidate prioritization. Standardized sample preparation, separation of native urinary constituents from processing-derived products, and validation in independent cohorts remain necessary.
\end{abstract}


\medskip
\noindent\textbf{Correspondence:} M. F. Pereira: \href{mailto:mauro.pereira@ku.ac.ae}{mauro.pereira@ku.ac.ae}

\medskip
\noindent\textbf{Abbreviations:} BPH, benign prostatic hyperplasia; BWO, backward-wave oscillator; GC--MS, gas chromatography--mass spectrometry; NMR, nuclear magnetic resonance; PC, prostate cancer; PHI, Prostate Health Index; PSA, prostate-specific antigen; THz, terahertz; VOC, volatile organic compound.
\endgroup

\twocolumn
\section{Introduction}

Prostate cancer (PC) is one of the most common malignancies in men \cite{ref1_RecentGlobalPatterns,ref2_IarcGlobalCancer,ref3_RussianCancerReport}. Despite significant advances in its diagnosis and treatment, the challenge of early detection remains crucial \cite{ref4_ProstateCancerUpdate}. The primary serum marker for prostate cancer, widely used for screening, is the prostate-specific antigen (PSA). PSA is a serine protease regulated by androgens and produced by both normal epithelial cells and prostate cancer cells. The analysis of PSA content in blood plasma is a simple, reliable, and inexpensive method to detect prostate cancer and is widely used \cite{ref5_BiologyOfProstate}. However, non-cancerous prostatic conditions can also result in elevated PSA levels. Conditions such as urinary tract infection (UTI), benign prostatic hyperplasia (BPH), and prostatitis can lead to elevated PSA levels \cite{ref6_DifferenDiagnosiOf}. Therefore, a high PSA level is not a conclusive indicator of the presence of a malignant prostate tumor. The Prostate Health Index (PHI) test is more effective than the traditional total PSA test, as it also detects free PSA and [-2]proPSA \cite{ref7_DiagnostAbilityOf}, leading to a more refined risk assessment. This allows for a more effective differentiation between prostate cancer and other conditions, such as BPH and prostatitis, compared to the PSA test. Another valuable marker is prostate cancer antigen 3 (PCA3). PCA3 expression refers to the level at which the non-coding RNA of prostate cancer antigen 3 (PCA3) is transcribed and present in cells, which is significantly higher in prostate cancer compared to normal or benign prostate tissue. PCA3 expression is detected in urine sediment after digital rectal massage, in which prostate cells exfoliate into the urethra \cite{ref8_MoleculaPca3Diagnost,ref9_Pca3Tmprss2Erg}.

The diagnostic importance of determining microRNA expression is being investigated. MicroRNAs are small non-coding RNAs about 22 nucleotides long that participate in the post-transcriptional regulation of gene expression. The potential for determining the expression of certain microRNAs in urine as markers of prostate cancer has been demonstrated \cite{ref10_Mir888Is,ref11_PotentiaUrinaryMirna,ref12_TheUtilityOf,ref13_DiagnostAndPrognost,ref14_AssessmeOfMir}.
\nocite{youden1950_index}

These tests refine the probability of clinically significant cancer, but do not provide histopathological confirmation. The examination of biopsy tissue remains the diagnostic reference standard; multiparametric magnetic resonance imaging can guide targeted sampling, although sampling error remains possible.

More research is needed to develop non-invasive tests based on metabolomics. Metabolites and their concentrations are directly related to the underlying biochemical activity and state of cells, tissues, and organs, providing an opportunity for the development of new diagnostic techniques. In particular, analysis of the chemical composition of exhaled breath and biological fluids (blood, saliva, and urine) can provide information about diseases and pathological processes \cite{ref15_TheHumanUrine,ref16_ImpactOfExercise}. Urine is an attractive sampling matrix because collection is simple and inexpensive, and many metabolites are present at measurable concentrations.

Existing techniques used in clinics to monitor metabolites are limited to standard laboratory methods, such as biochemical blood tests and clinical urine analysis. More advanced academic studies are performed mainly using chromatography or mass spectrometry \cite{ref17_SelectedIonFlow,ref19_TheUntargetUrine}. Ref.~\cite{ref17_SelectedIonFlow} reviews mass spectrometry for trace-gas analysis in medicine and the environment. A database of specific urine VOCs (urine volatilome), consisting of 841 compounds from 80 different chemical classes, including substances reported in oncological settings, is provided in Ref.~\cite{ref19_TheUntargetUrine}.

A statistical analysis of data obtained from the examination of urine samples using 1H-nuclear magnetic resonance (1H-NMR) led to the identification of potential biomarkers of prostate cancer of 20 detected metabolites in Ref.~\cite{ref20_NovelMetaboliSignatur}. Urine metabolites were profiled using mass spectrometry and multifactorial statistical analysis to predict the outcomes of prostate cancer treatment in~\cite{ref21_UntargetUrineMetaboli}. This study revealed that the presence of fragments of the monoacylglycerin skeleton of glycerides can statistically correlate with the progression of the disease.

Terahertz (THz) and mid-infrared (MIR) photonics are developing rapidly with the increasing availability of new materials, sources, and detectors \cite{ref22_The2017Terahert,ref23_TeraMirRadiatio,10.1021/acsomega.3c10175,Cousin2022}. The strong vibrational and rotational resonances of target molecules in these ranges can potentially turn metabolomics into one of their most impactful applications, but a relatively small number of metabolic biomarkers have concrete THz signatures assigned.

{High-resolution terahertz gas spectroscopy is a sensitive spectral-analysis technique for detecting chemical compounds in multicomponent gas mixtures}, including trace {components}. It combines high sensitivity and resolution. Specific spectral ranges can be identified in which the spectral lines of target molecules do not overlap, allowing the measurements of complex combinations of components in gas mixtures \cite{ref24_ApplicatOfMicrowav}.


{
High-resolution THz spectroscopy has previously been used to characterize urine thermal-decomposition products in cancer patients undergoing chemotherapy \cite{ref25_SensingNitrilesWith}. Vaks et al. subsequently applied this approach to investigate the molecular composition of urine-derived products from patients with PC \cite{ref26_ApplicatOfHigh}. Later reports extended the investigation to comparisons between PC and BPH \cite{ref44_Vaks2024IEEE,ref45_Vaks2025JRE,vaks2025studying}. The scope and available data of these reports are summarized in Table S3 of the Supporting Information. These related studies represent successive developments within the same research programme and provide the experimental and methodological background for the present review and updated analysis.}

This article combines a critical narrative review of high-resolution THz urine spectroscopy with a statistical reanalysis of the 24-PC / 14-BPH clinical dataset (summarized in Tables 1 and 2) and a comparative interpretation of the associated molecular assignments. Complete molecular-transition tables supporting the spectroscopic assignments are provided in the Supporting Information (Tables S1 and S2).
 
 Statistical reanalysis of serum PSA uses nonparametric group comparison and receiver-operating-characteristic analysis with bootstrap uncertainty estimation to assess between-group differences, patient-level overlap and the uncertainty in PSA-based discrimination. The spectroscopic analysis examines molecular assignments and representative spectral features to distinguish compounds shared between PC and BPH from candidate PC-associated volatile and thermal-decomposition products.
A central objective is to relate these THz assignments to urinary metabolomic findings obtained using other analytical methods, including carbonyl profiling, gas chromatography–mass spectrometry and proton nuclear magnetic resonance spectroscopy \cite{ref19_TheUntargetUrine,ref20_NovelMetaboliSignatur,lin2021_quantSchema, ref44_MechanisAndConseque}. This comparison distinguishes reports of the same compounds from broader relationships involving chemical classes or metabolic pathways, while recognizing that chemical correspondence does not by itself establish PC specificity. Therefore, particular attention is given to the distinction between native urinary constituents and products generated during sample processing. By integrating the clinical statistical analysis, and comparative chemical interpretation, this review aims to prioritize molecular candidates and identify the requirements for quantitative, prospective evaluation of THz urine spectroscopy.

\section{Materials and Methods}

\subsection{Review scope, evidence levels, and secondary analysis}

{This article combines a critical narrative review of high-resolution THz urine spectroscopy with statistical reanalysis of the 24-PC/14-BPH clinical dataset (summarized in Tables 1 and 2) and comparative interpretation of the associated molecular assignments. The review considers the technical basis of the spectroscopic method \cite{ref27_HighResolutiTerahert,ref28_OnThePossibil,ref29_ApplicatOfA}, its application to urine-derived products in cancer patients and comparisons between PC and BPH \cite{ref25_SensingNitrilesWith,ref26_ApplicatOfHigh,ref44_Vaks2024IEEE,ref45_Vaks2025JRE} and selected urinary metabolomics studies that provide complementary chemical and metabolic context \cite{ref19_TheUntargetUrine,ref20_NovelMetaboliSignatur,lin2021_quantSchema,ref42_IdentifiOfA,ref46_NonInvasiveUrinary,ref47_1hNmrBased}.
The reviewed sources were assessed in terms of cohort definition, sample collection and processing, the type and availability of numerical data, and the conclusions supported by those data. Clinical metadata, patient-indexed spectra, patient-by-molecule feature matrices, transition-assignment catalogues and group-level summaries were considered separately. This distinction was used to differentiate evidence supporting molecular assignments and exploratory comparisons from evidence supporting patient-level diagnostic discrimination. The 24-PC/14-BPH clinical table was transcribed for a PSA reanalysis. The empirical receiver-operating-characteristic area under the curve (AUC) was accompanied by a 20,000-resample nonparametric bootstrap interval. The threshold maximizing Youden's index was treated as a descriptive, cohort-specific operating point \cite{youden1950_index}. PSA distributions in the PC and BPH groups were compared using a two-sided Mann–Whitney U test.}

\begin{figure*}
    \centering
    \includegraphics[width=1\linewidth]{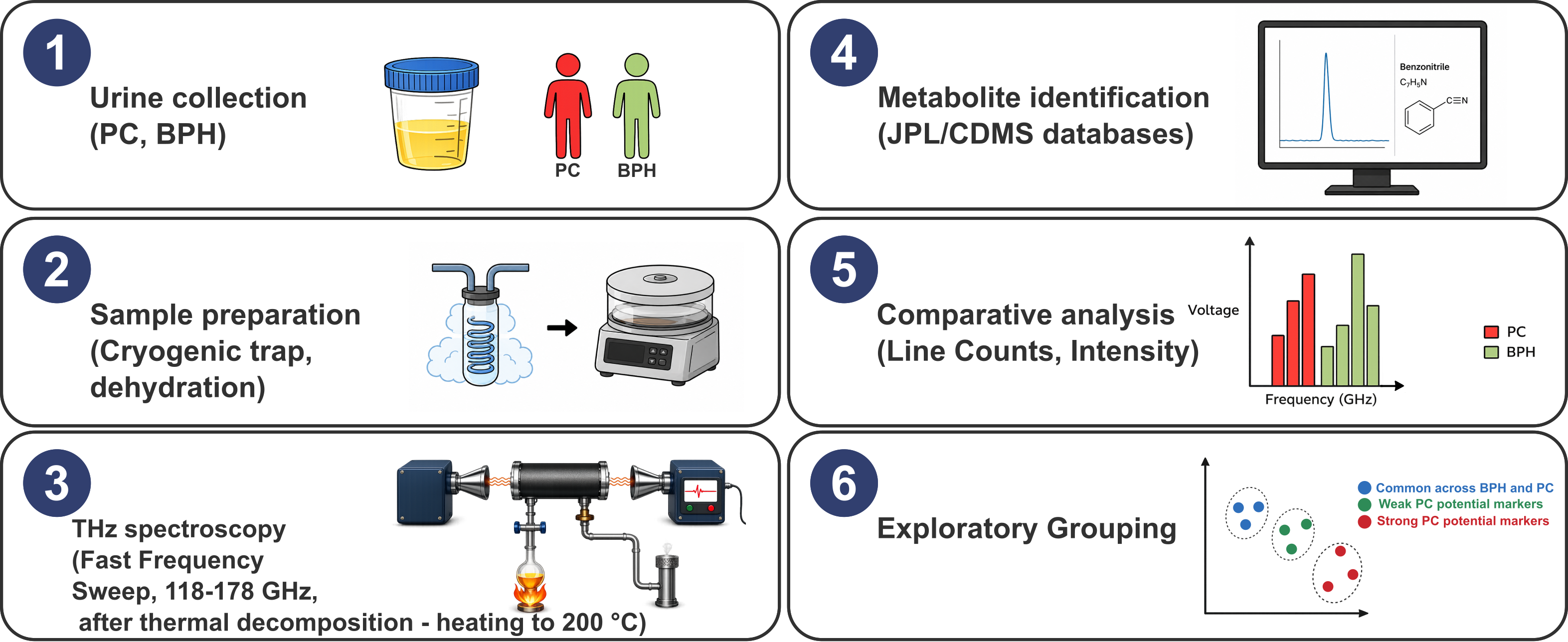}
    \caption{Clinical and analytical workflow for the recent PC/BPH dataset, placed in the context of the existing THz urine-spectroscopy literature.
The workflow follows urine collection, vacuum drying to form a thin film, fast-sweep THz gas spectroscopy, molecular
assignment using the JPL catalogue, comparative line-count analysis and exploratory grouping for candidate prioritization.}
    \label{fig:workflow}
\end{figure*}

Figure \ref{fig:workflow} gives a schematic representation of the study.

Urine samples were taken from patients with a verified diagnosis of prostate cancer (PC) and benign prostatic hyperplasia (BPH). All patients provided their informed written consent to collect biological fluid samples. A total of 24 urine samples from prostate cancer patients were studied. The mean age was 64.5 {years (range 49--81)}. The mean preoperative PSA value was 14.5 ng mL$^{-1}$ (range 4.6--47.32 ng mL$^{-1}$). All patients underwent radical prostatectomy with extended lymphadenectomy. Urine sampling was performed in the morning prior to the operation. After collection, the urine samples were frozen for further spectroscopic study. Data for PC patients are presented in Table~\ref{tab:pc-patients}.

The comparison group consisted of 14 patients with benign prostatic hyperplasia (BPH). This diagnosis was confirmed by histological examination of material obtained after transurethral resection of the prostate. The mean age was 66.9 years (range 58--83), and the mean preoperative PSA value was 6.5 ng mL$^{-1}$ (range 1.4--19 ng mL$^{-1}$). Urine sampling was performed in the morning before surgery after prostate massage. Data on patients with BPH are presented in Table~\ref{tab:bph-patients}. A comparison of substances in urine from patients with PC and BPH is provided in Table~\ref{tab:substance-comparison}.
\\
\begin{figure*}
    \centering
    \includegraphics[width=0.78\textwidth]{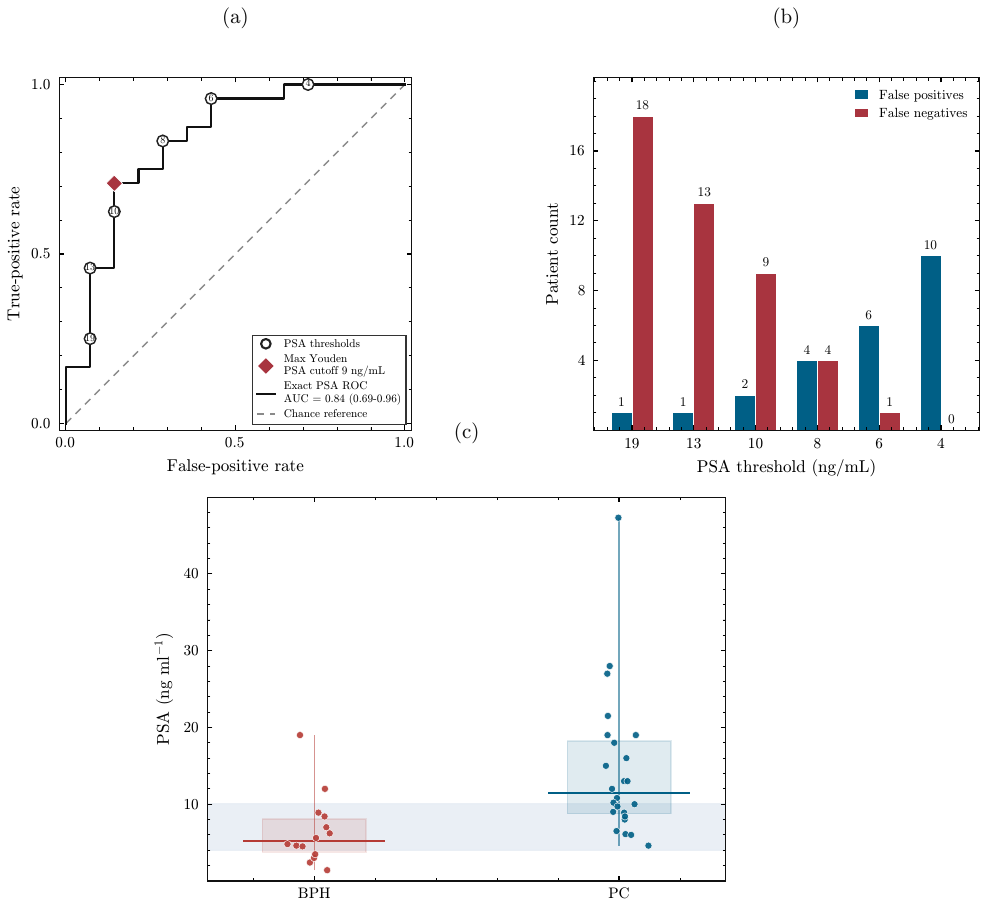}
    \caption{Threshold-based diagnostic analysis of serum PSA for distinguishing prostate cancer (PC) from benign prostatic hyperplasia (BPH) in the study cohort:
(a) Empirical receiver-operating-characteristic (ROC) curve for PSA-based separation of PC patients ($n=24$) from BPH patients ($n=14$). Open circles indicate selected PSA thresholds, and the red diamond marks the maximum-Youden operating point \cite{youden1950_index}. The dashed line represents chance-level discrimination. (b) False-positive and false-negative patient counts obtained at selected PSA thresholds. Blue bars indicate BPH patients incorrectly classified as PSA-positive, and red bars indicate PC patients incorrectly classified as PSA-negative. (c) Patient-level PSA distributions in the BPH and PC groups. Points represent individual patients; horizontal bars indicate group medians, shaded boxes indicate interquartile ranges, and vertical lines show the observed ranges. The pale blue band marks the clinically ambiguous PSA range of 4--10 ng mL$^{-1}$.}
    \label{fig:psa-performance}
\end{figure*}

{Figure~\ref{fig:psa-performance} evaluates how much diagnostic information is provided by serum PSA alone in this cohort. This was not a trained predictive model; PSA was analysed as a single continuous diagnostic variable by applying possible thresholds to the observed patient values. PC patients had a median PSA of 11.4 ng mL$^{-1}$ and BPH patients a median of 5.2 ng mL$^{-1}$ (Mann--Whitney $U=283$, two-sided $p=0.00053$). The empirical AUC was 0.842, with a 20,000-resample bootstrap 95\% interval of 0.692-0.960. The maximum-Youden threshold was $\geq$ 9.0 ng mL$^{-1}$, giving an apparent sensitivity of 70.8\% and specificity of 85.7\% \cite{youden1950_index}. The threshold was selected and evaluated on the same small cohort, so it is an in-sample descriptor rather than a universal clinical cutoff. The patient-level overlap supports the need for complementary molecular information rather than reliance on PSA alone.}

\subsection{THz frequency sweeping spectrometer}
A number of spectrometers based on non-stationary effects have been developed and implemented at IPM RAS, utilising various types of radiation modulation techniques \cite{ref27_HighResolutiTerahert,ref28_OnThePossibil,ref29_ApplicatOfA}. For example, a spectrometer based on phase switching has recently been employed in the terahertz range to detect nitriles in urine vapours from cancer patients undergoing chemotherapy \cite{ref25_SensingNitrilesWith}. Such spectrometers enable sweeping over wide frequency ranges, including through harmonic generation in semiconductor superlattices \cite{ref30_GiantControllGigahert,ref31_AnalyticExpressiFor,ref32_HarmonicGeneratiIn,ref33_TerahertGeneratiBy,Oriaku:17,ApostolakisPereira+2020+3941+3952,10.1117/1.JNP.13.036017,ref34_ProgressInAnalytic,10.1063/1.5050917}.

\begin{figure}
    \centering
    \includegraphics[width=.94\linewidth]{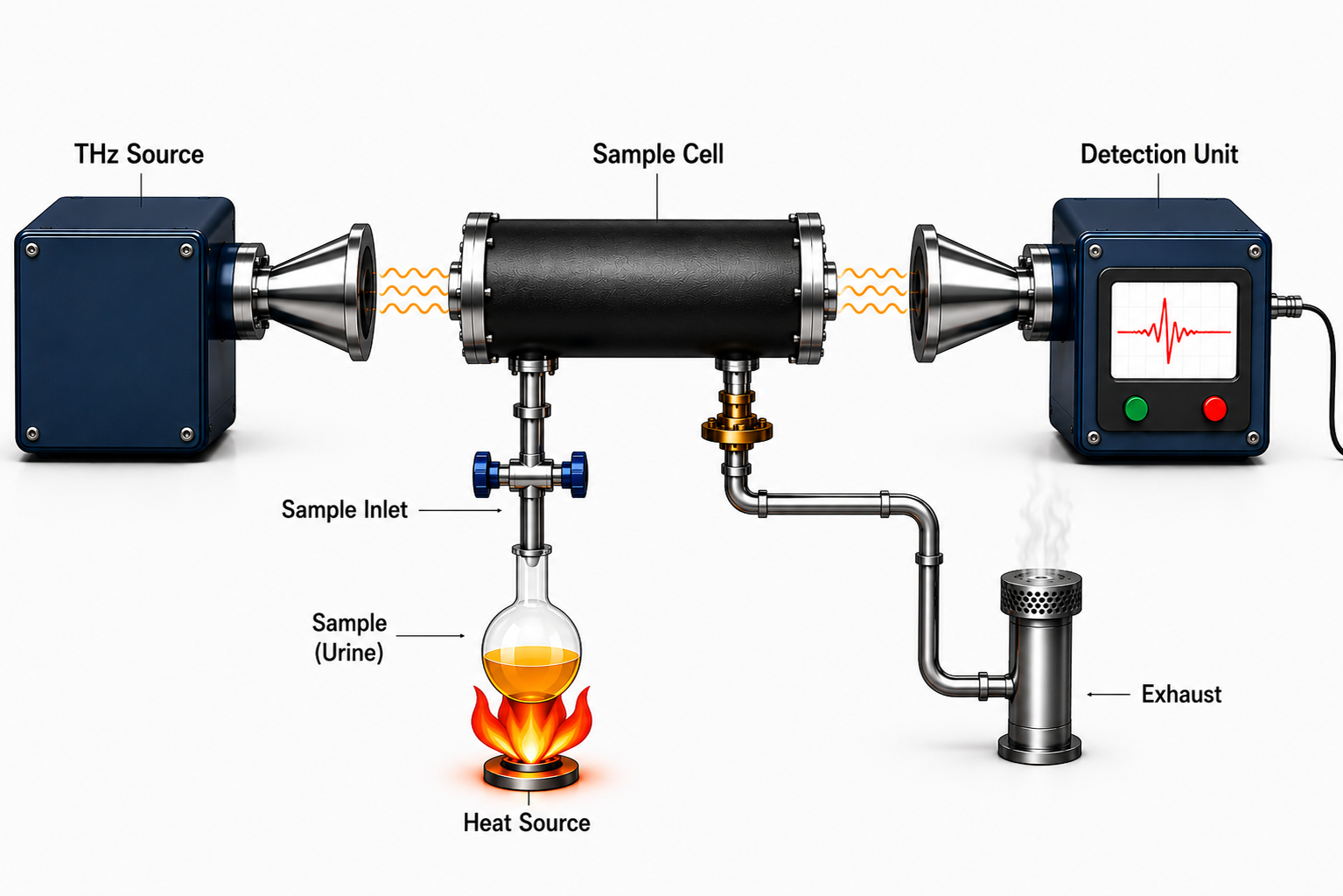}
    \caption{Fast-sweep THz spectrometer schematic and measurement principle used to detect spectra of urine thermal-decomposition products.}
    \label{fig:setup}
\end{figure}

In the present manuscript, we employ an enhanced fast frequency-sweeping spectrometer operating in a non-stationary detection regime. While the general operating principle of 2 mm-range spectrometers with fast frequency sweep is similar, the accessible frequency range and stability of each system are influenced by the characteristics of the radiation source used, in particular the backward wave oscillator (BWO). Depending on the operating region, BWOs may exhibit variations in output stability, especially near the edges of the tuning range, and the effective sweep range may therefore be slightly adjusted in practice. The BWO-based fast frequency-sweep spectrometer used in this study operates in the 118--175 GHz range. Figure~\ref{fig:setup} illustrates the device.

In fast frequency-sweep operation, the radiation source is linearly frequency-modulated and rapidly swept across molecular resonances. When the instantaneous frequency of the source crosses a molecular transition, a macroscopic molecular polarization is excited, which subsequently decays with a characteristic relaxation time governed by collisional and Doppler broadening mechanisms. The detected signal arises from the interference between the transmitted frequency-chirped radiation and the free-decaying molecular polarization. As a result, the spectrometer records a transient time-domain response consisting of a sharp resonance marker followed by a damped oscillatory signal. In the fast-sweep limit, the recorded transient differs from a stationary absorption profile. Instead, spectral information is encoded in the temporal position and structure of the transient response and is converted into frequency using the known sweep law of the source. Depending on the relative phase of the interfering fields and the polarity of the detection chain, the observed feature may appear either as a dip (see the representative dips in Fig.~\ref{fig:spectra-snr}) or as an inverted signal. The physical basis of fast frequency-sweep spectroscopy, including the distinction between slow- and fast-passage regimes and the role of molecular relaxation processes, has been analysed in detail by Vaks et al.~\cite{ref28_OnThePossibil}.

Urine is a biological fluid that contains nitrogenous products from the decomposition of protein substances, such as urea, uric acid, hippuric acid, creatinine, xanthine, urobilin, indican, and various salts, primarily chlorides, sulfates, and phosphates \cite{ref15_TheHumanUrine}. Additionally, it may contain amino acids, residual amounts of glucose, and protein. These substances collectively form the primary metabolome of urine. In cases of inflammatory, urological, and oncological diseases, the composition of urine can change, leading to increased concentrations of proteins, glucose, and ketone bodies \cite{ref15_TheHumanUrine}. All these chemical compounds are considered primary metabolites found in urine. However, due to their low volatility, detecting them using gas THz spectroscopy is challenging. Therefore, the method for analyzing urine samples involves heating and thermal decomposition to obtain vaporous substances and gaseous degradation products of primary metabolites.

When heated, substances in urine undergo thermal decomposition, forming volatile products that can be detected by terahertz spectroscopy \cite{ref24_ApplicatOfMicrowav,ref25_SensingNitrilesWith}. The frequency-sweeping, high-resolution gas spectrometer measured the 118--175 GHz range within 0.5 min, allowing characteristic substances to be detected at trace levels \cite{ref26_ApplicatOfHigh,ref27_HighResolutiTerahert}. During sample preparation, a 1--2 mL urine sample was thawed and transferred into a sterile, chemically clean flask. The sample was dehydrated by vacuum drying because excess water hinders trace-level spectroscopic measurements. After dehydration, a thin film formed in the sample flask. Volatile substances from the resulting residue were released into the measuring cell, first without heating and then with heating. The measuring cell had previously been evacuated to $10^{-3}$ mbar. Before introducing the thermal-decomposition products, the full spectrometer range was recorded without sample intake to characterize background absorption lines. Absorption lines were assigned using spectroscopic molecular databases \cite{ref35_JplSpectralCatalog}.

{For a qualitative assessment of sample-to-sample differences, the number of registered absorption lines assigned to each compound was calculated within the same spectrometer range under consistent measurement conditions. These line counts were used as assignment-level evidence for comparing molecular patterns between PC and BPH samples.}

{The resulting molecular assignments were then summarized according to their occurrence and prioritization across the PC and BPH cohorts. Compounds were grouped into three interpretation categories: common across BPH and PC, weak potential PC markers, and strong potential PC markers. This categorization was used to generate the molecule-map visualization and to prioritize candidate urine-derived volatile and thermal-decomposition products for future validation.}

\begin{table*}[!t]
\centering
\begingroup
\fontsize{6.9}{7.15}\selectfont
\renewcommand{\arraystretch}{0.89}
\captionof{table}{Information about prostate cancer (PC) patients. RBC = red blood cells; Hgb = hemoglobin; TNCT = too numerous to count.}\label{tab:pc-patients}
\begin{tabular}{@{}>{\raggedright\arraybackslash}p{0.08\textwidth}>{\raggedright\arraybackslash}p{0.06\textwidth}>{\raggedright\arraybackslash}p{0.16\textwidth}>{\raggedright\arraybackslash}p{0.12\textwidth}>{\raggedright\arraybackslash}p{0.06\textwidth}>{\raggedright\arraybackslash}p{0.07\textwidth}>{\raggedright\arraybackslash}p{0.27\textwidth}@{}}
\toprule
Sample No. & Age & Stage & Gleason score & ISUP & PSA & General urinalysis \\
\midrule
3 & 66 & pT3aN0M0 & 3+4=7 & 2 & 6.1 & Normal \\
4 & 70 & pT3bN1M0 & 3+4=7 & 2 & 27 & Normal \\
7 & 60 & pT2cN0M0 & 3+4=7 & 2 & 8.9 & Normal \\
8 & 59 & pT3N1M0 & 4 + 4 = 8 & 4 & 15 & Normal \\
9 & 76 & pT2cN0M0 & 3 + 3 = 6 & 1 & 6 & Glucose 6 mmole/L \\
13 & 62 & pT3bN0M0 & 3 + 4 = 7 & 2 & 47.32 & Normal \\
14 & 73 & pT2aN0M0 & 3 + 3 = 6 & 1 & 6.5 & RBC---100 cells/\textmu{}L \\
15 & 58 & pT3bN0M0 & 3 + 4 = 7 & 2 & 28 & Normal \\
16 & 58 & pT2cN0M0 & 3 + 4 = 7 & 2 & 19 & Normal \\
17 & 55 & pT2cN0M0 & 4 + 4 = 8 & 4 & 10.8 & Hgb---25 g/L \\
18 & 74 & pT2cN0M0 & 3 + 3 = 6 & 1 & 9 & RBC---12 cells/\textmu{}L \\
19 & 62 & pT3aN0M0 & 4+3=7 & 3 & 18 & Normal \\
21 & 54 & pT2N0M0 & 3+4=7 & 2 & 16 & TNCT \\
22 & 74 & pT3aN0M0 & 3+4=7 & 2 & 13 & Normal \\
23 & 59 & pT3aN0M0 & 4+5=9 & 5 & 8 & Normal \\
24 & 76 & pT2cN0M0 & 3+4=7 & 2 & 8.4 & Normal \\
26 & 66 & pT2cN0M0 & 3+4=7 & 2 & 4.6 & Normal \\
27 & 64 & pT3bN1M0 & 4+4=8 & 4 & 10.2 & Normal \\
28 & 56 & pT3bN0M0 & 3+3=6 & 1 & 12 & Normal \\
29 & 72 & pT3bN1aM1a & 4+4=8 & 4 & 21.5 & Normal \\
30 & 58 & pT3bN0M0 & 4+4=8 & 4 & 13 & Normal \\
32 & 49 & pT2cN0M0 & 3+3=6 & 1 & 10 & Normal \\
33 & 81 & pT2cN0M0 & 3+4=7 & 2 & 9.7 & Normal \\
46 & 65 & pT2cN0M0 & 3+4=7 & 2 & 19 & Normal \\
\bottomrule
\end{tabular}
\par\vspace{2pt}

\captionof{table}{Information about benign prostatic hyperplasia (BPH) patients. Le = leukocyte esterase.}\label{tab:bph-patients}
\begin{tabular}{@{}>{\raggedright\arraybackslash}p{0.14\textwidth}>{\raggedright\arraybackslash}p{0.14\textwidth}>{\raggedright\arraybackslash}p{0.14\textwidth}>{\raggedright\arraybackslash}p{0.44\textwidth}@{}}
\toprule
Sample No. & Age & PSA & General urinalysis \\
\midrule
25 & 83 & 8.9 & Le---15 \\
31 & 60 & 19 & Normal \\
34 & 65 & 8.4 & Normal \\
35 & 60 & 1.4 & Normal \\
36 & 73 & 4.8 & Normal \\
37 & 62 & 4.6 & Normal \\
38 & 60 & 5.6 & Normal \\
39 & 70 & 2.4 & Normal \\
40 & 74 & 3 & Normal \\
41 & 69 & 4.5 & Normal \\
42 & 58 & 7 & Normal \\
43 & 67 & 12 & Normal \\
44 & 64 & 3.48 & Normal \\
45 & 71 & 6.2 & Normal \\
\bottomrule
\end{tabular}
\par\vspace{2pt}

\captionof{table}{Comparison of substances in urine from PC and BPH patients.}\label{tab:substance-comparison}
\begin{tabular}{@{}>{\raggedright\arraybackslash}p{0.46\textwidth}>{\raggedright\arraybackslash}p{0.46\textwidth}@{}}
\toprule
Substances that are present in both PC and BPH samples & Substances that are present in PC samples but are absent in BPH samples \\
\midrule
1. Isocyanic acid 
2. Acetic acid
3. Propanediol
4. Acetaldehyde
5. Acetonitrile
6. Methylbutyronitrile (its relative concentration in PC samples is higher than in BPH samples)
7. Pentadiene nitrile (its relative concentration in PC samples is higher than in BPH samples)
8. Ethynyl benzonitrile (its relative concentration in PC samples is higher than in BPH ones)
9. Methyl mercaptan (its relative concentration in PC samples is higher than in BPH ones)
10. Alanine
11. Urea (its relative concentration in PC samples is higher than in BPH samples)
12.  Ethylene glycol
13. Methyl carbamate
14. Cyanoethynylbenzene
15. Carbonyl sulphide
16.  Sulphur dioxide & 1. Formic acid
2. Phenol
3. Methylphenyl ether
4. Propanal
5. Benzaldehyde
6. Glycolaldehyde
7. Malondialdehyde
8. Butyronitrile
9. Pentanenitrile
10. Methyl isocyanate \\
\bottomrule
\end{tabular}
\endgroup
\end{table*}

\section{Results and Discussion}

The results obtained from high-resolution THz spectroscopy of urine samples from patients with prostate cancer (PC) and benign prostatic hyperplasia (BPH) reveal {differences in urine-derived volatile and thermal decomposition product assignments}. In this section, we analyze these differences in detail, discuss their possible biochemical origins, and compare them with existing metabolomic findings reported in the literature.

\begin{figure*}[!t]
\centering
\includegraphics[width=0.90\textwidth]{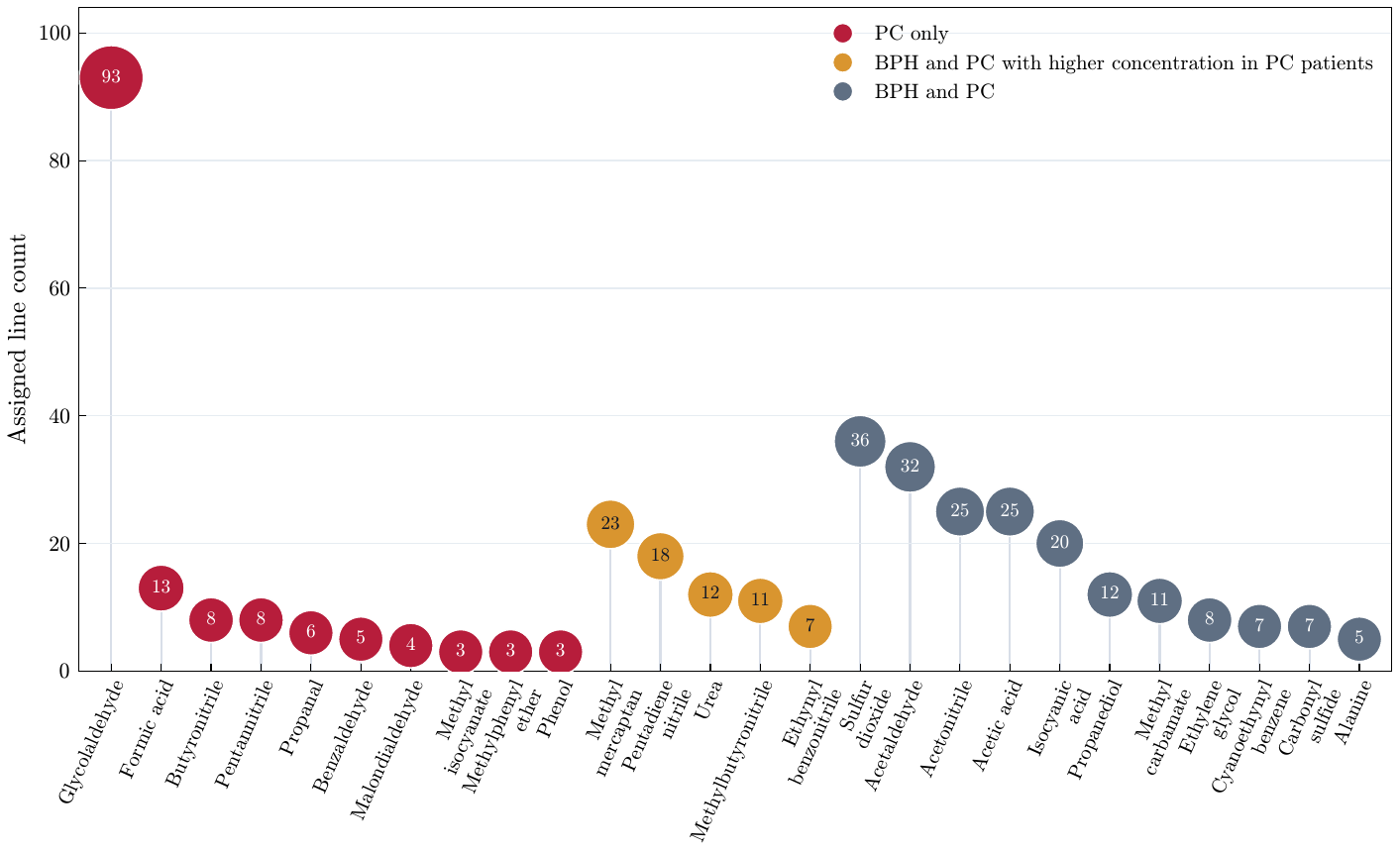}
\caption{Molecular assignment landscape of urine-derived volatile products detected in the BPH and PC cohorts. Colours indicate PC-only assignments, compounds detected in both BPH and PC, and compounds detected in both groups with higher apparent abundance in PC patients. Lollipop height, marker size and marker number report the assigned rotational-transition count from Tables S1 and S2 in the Supporting Information.}
\label{fig:molecular-landscape}
\end{figure*}

Several substances (Table~\ref{tab:substance-comparison}) were present in all samples, regardless of the type of prostate pathology. These include isocyanic acid, urea, acetic acid, and others. To interpret the detected volatile products, the molecular assignments were organized into three practical categories: compounds assigned only in PC samples, compounds detected in both BPH and PC, and compounds detected in both groups but with stronger evidence or higher apparent abundance in the PC group (Fig.~\ref{fig:molecular-landscape}).

Within this assignment-table summary, glycolaldehyde had the largest number of catalogue-matched transitions among the PC-only category, followed by formic acid, butyronitrile, pentannitrile, propanal, benzaldehyde, malondialdehyde, methyl isocyanate, methylphenyl ether and phenol. These compounds should be interpreted as candidate PC-associated urine-derived volatile or thermal-decomposition products that require targeted validation, rather than as established clinical biomarkers. The BPH-and-PC category contains shared prostate-pathology or sample-processing chemistry, whereas the PC-enriched shared category highlights compounds detected in both cohorts but more prominent in PC samples, including methyl mercaptan, pentadiene nitrile, urea, methylbutyronitrile and ethynyl benzonitrile.


{Figure~\ref{fig:molecular-landscape} summarizes the molecular-assignment inventory. For each molecule, the number printed at the marker and the lollipop height represent the number of catalogue-matched rotational transitions listed in Tables S1 and S2 in the Supporting Information. These values do not represent molecular concentration or the magnitude of the between-group difference. The colours indicate the classification derived from the sample-level comparison described in the Methods: detected only in PC samples, detected in both BPH and PC samples, or detected in both groups with greater apparent abundance in the PC samples. }

\begin{figure*}[!htp]
\centering
\includegraphics[width=0.86\textwidth]{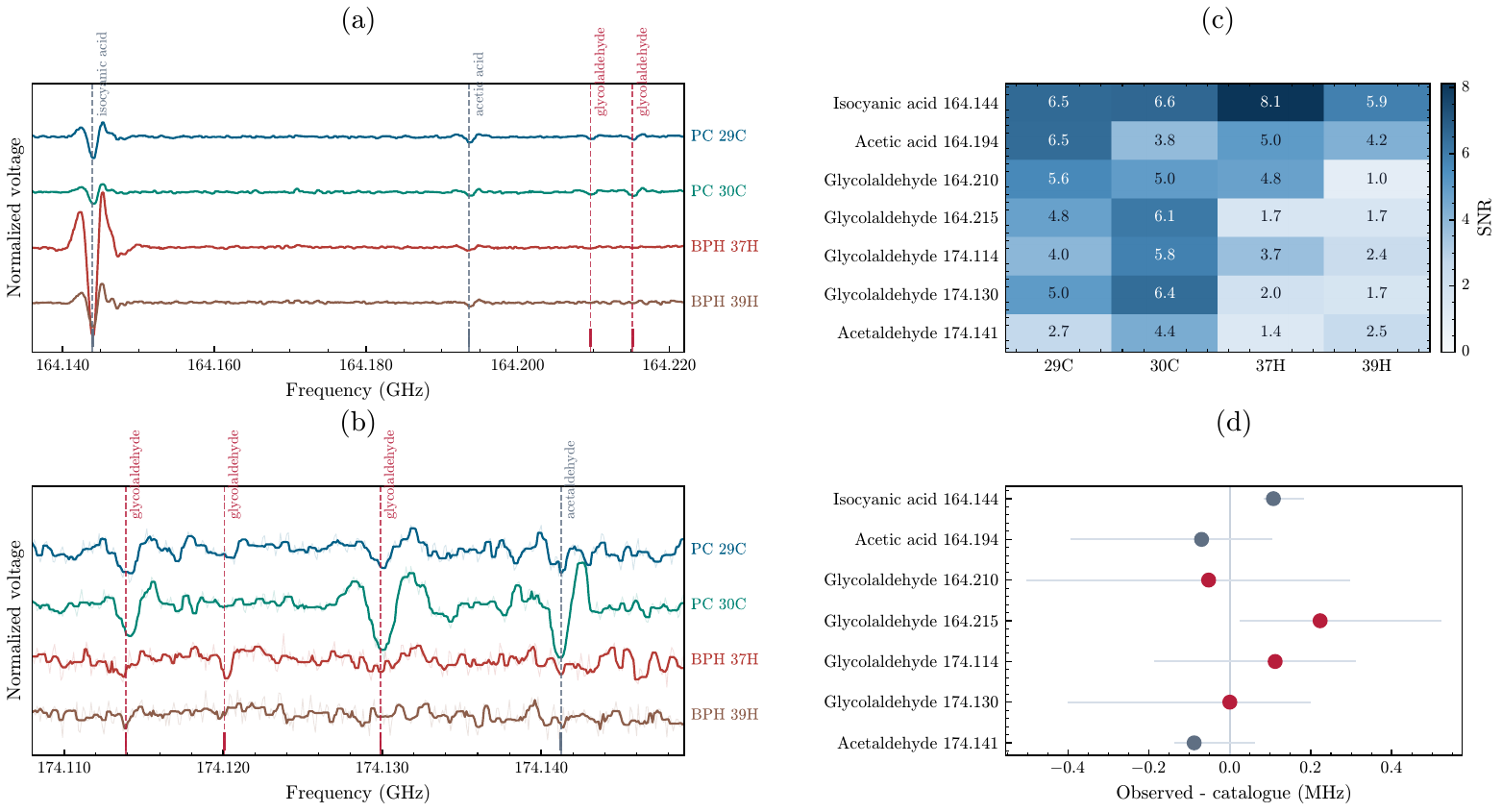}
\caption{(a) Normalized voltage spectra from PC and BPH urine-derived volatile products in the 164 GHz window. Dashed lines mark catalogue transition frequencies. (b) Normalized spectra from the same samples in the 174 GHz window, highlighting additional glycolaldehyde transitions and an acetaldehyde comparison line. (c) Local signal-to-noise ratio matrix for the marked transitions. Numbers inside the squares report the signal-to-noise ratio for each molecule-frequency assignment in each sample; darker colour indicates stronger local detectability. (d) Difference between observed and catalogue frequency residuals for the same assignments. Values closer to zero indicate better agreement between measured spectral features and catalogue frequencies.}
\label{fig:spectra-snr}
\end{figure*}

Figure~\ref{fig:spectra-snr} shows representative detector-voltage traces in the 164 and 174 GHz spectral windows together with the corresponding catalogue assignments.

Spectra from two benign prostatic hyperplasia (BPH) patients (37H and 39H) are compared with spectra from two prostate cancer (PC) patients (29C and 30C). Isocyanic acid (HNCO) and acetic acid (CH$_{3}$CO$_{2}$H) are detected in both BPH and PC samples in the selected windows. Glycolaldehyde (C$_{2}$H$_{4}$O$_{2}$) is consistently detected in the representative PC spectra and is weaker or absent in the paired BPH traces at the marked transitions; acetaldehyde (C$_{2}$H$_{4}$O) is shown as an additional carbonyl comparison line in the 174 GHz window.

In the 164 GHz region, multiple closely spaced transitions of isocyanic acid are not spectrally resolved and are therefore recorded as an envelope centred at 164.1439173 GHz. A similar unresolved envelope is observed for acetic acid, comprising nearby transitions with a central frequency of 164.1935948 GHz. In both PC samples, glycolaldehyde exhibits distinct signatures at 164.2096024 GHz and 164.2151767 GHz. Additional glycolaldehyde features are observed in the 174 GHz region at 174.1138879 GHz and 174.1299 GHz in both PC samples. The SNR matrix in Fig.~\ref{fig:spectra-snr} makes this visual comparison explicit, while the residual panel shows that the marked features remain close to catalogue frequencies.

These representative spectra support the molecular assignments and highlight candidates for further investigation. Their diagnostic potential will require quantitative evaluation using cohort-wide intensity data and independent validation.

\begin{figure}[!htp]
\centering
\includegraphics[width=\linewidth,keepaspectratio]{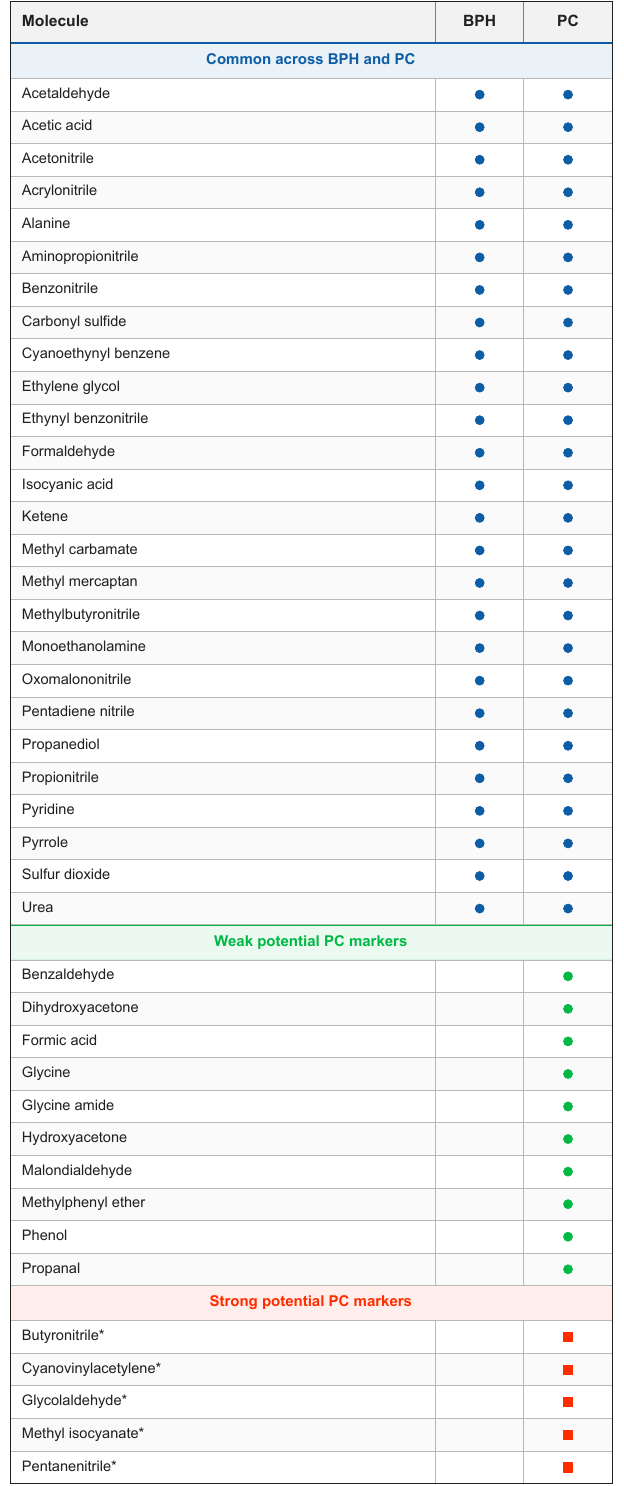}
\caption{BPH/PC molecule map separating shared chemistry from weak and strong candidate PC-associated features. Section classifications are presented as table headers, with BPH and PC shown in the same reading orientation as the molecule names. All candidates remain validation targets rather than confirmed diagnostic markers.}
\label{fig:dendrogram-replacement}
\end{figure}

The interpretation of {candidate-prioritization} is summarized in the consolidated molecule map in Fig. \ref{fig:dendrogram-replacement}. This map integrates the evidence from the assignment-table, the exploration grouping {and the literature audit} into a single marker-prioritization view. Compounds shared between BPH and PC are treated as prostate-pathology or sample-processing chemistry rather than cancer-specific markers. Candidate PC-associated compounds are retained as validation targets only when they are supported by assignment tables, {exploratory grouping}, or evidence from the literature.

This prioritization emphasizes that acetonitrile, sulfur dioxide, and other BPH/PC-shared compounds should not be claimed as PC-specific. In contrast, molecules such as glycolaldehyde, butyronitrile, pentanenitrile, and methyl isocyanate remain useful follow-up candidates because they appear in the PC-associated assignment or {exploratory grouping}. Table~\ref{tab:weak-strong-pc-markers} separates molecules found in PC samples only {into stronger follow-up candidates and weaker contextual signals based on the relevance of the molecules in the literature}.

The basis of metabolic analysis in the search for markers that distinguish malignant tumors from benign tissues is the altered biochemical activity of pathological cells and their surrounding tissue environment. In this pilot study, THz spectroscopy of urine-derived volatile products revealed differences between PC and BPH in both assignment-table evidence and representative spectral windows. These differences should be interpreted as exploratory marker-prioritization evidence rather than as validated diagnostic performance.

At this point, it is useful to compare our findings with those in the existing literature. The PC-enriched shared group includes methylbutyronitrile, pentadiene nitrile, ethynyl benzonitrile, and methylmercaptan, which were detected in both PC and BPH samples but were more prominent in PC on the basis of the assignment-table evidence. This observation is consistent with studies suggesting that the thermal decomposition of urinary amino-acid or protein-derived material can produce volatile products containing nitrile \cite{ref37_UrinaryMetaboliBiomarke}. Sulfur-containing compounds, including sulfur dioxide and methyl mercaptan, are best interpreted as shared prostate-pathology or processing-related chemistry unless cohort-level intensity data establish PC specificity \cite{ref26_ApplicatOfHigh}.

These findings are consistent with studies of 1H-nuclear magnetic resonance (1H-NMR) that have highlighted guanidinoacetate, phenylacetylglycine and glycine as potential urinary markers for prostate cancer \cite{ref20_NovelMetaboliSignatur}. In the THz assignment landscape reported here, several carbonyl compounds, including benzaldehyde, glycolaldehyde, and malondialdehyde, fall within the PC-associated or PC-enriched interpretive groups. Glycolaldehyde is particularly prominent in the assignment-table lollipop and in the representative spectral windows. It is a reactive carbonyl metabolite that can be linked to glycolytic or oxidative chemistry \cite{ref38_GlycolalAReactive}, and independent carbonyl metabolite profiling has detected glycolaldehyde in human urine in a clinical setting of risk for cancer \cite{lin2021_quantSchema}. Given the Warburg effect and the altered redox status in tumors \cite{ref39_TheWarburgEffect}, glycolaldehyde is a plausible validation target, but its native urinary origin and disease specificity still require targeted confirmation.

Malondialdehyde (MDA) is formed as a result of the peroxidation of polyunsaturated fatty acids under the action of reactive oxygen species (ROS) and is a marker of oxidative stress \cite{ref40_SerumMalondiaMda,ref53_UrinaryMalondiaMda}. Because oxidative stress and inflammation can also occur in benign prostate disease, MDA should be interpreted as a supportive biological context rather than as a PC-specific marker alone.

In microbial systems, intermediates formed during aromatic amino-acid metabolism, including pathways derived from phenylalanine, can cause benzaldehyde \cite{ref41_PhenylalanineBenzaldehyde}; this source supports the chemical pathway, but does not establish a specific origin of the human urinary or prostate. In particular, other research employing gas chromatography-mass spectrometry (GC-MS) has also identified aldehydes, such as 3-phenylpropionaldehyde and 2,5-dimethylbenzaldehyde, as significant markers of the disease \cite{ref42_IdentifiOfA}. A schematic comparison of the techniques is given in Table~\ref{tab:diagnostic-techniques}.

{Recent studies extend the independent evidence for prostate-disease-associated urinary metabolic patterns. Tang et al. used untargeted UPLC-MS/MS in
50 PC patients, 50 patients with histologically confirmed BPH and
30 participants without prostate disease, identifying 79 metabolites
contributing to PC--BPH discrimination, including lipid, amino-acid,
dipeptide and androgenic-steroid signals \cite{Tang2025UrineMetabolomics}. In 201 men undergoing prostate biopsy, Panach-Navarrete et al. combined 28 post-prostate-massage urinary $^{1}$H~NMR metabolites with prostate volume and ultrasound findings,
reporting an AUC of 0.89; however, the comparator comprised biopsy-negative
patients rather than a specifically defined BPH cohort
\cite{PanachNavarrete2026UrineNMR}. Complementary approaches include paper spray ionization mass spectrometry
(PSI-MS), for which Mahmud et al. reported separation of healthy, BPH,
prostatitis and PC groups in a 40-sample pilot study
\cite{Mahmud2026PaperSpray}, and olfactory-receptor nanodisc sensing,
for which Yoo et al. reported PC-associated urinary response patterns
and a cross-validated AUC of 0.964 in an expanded 290-sample dataset
\cite{Yoo2026OlfactoryNanodiscs}. The latter study's healthy-control
comparison does not establish PC--BPH discrimination. Collectively,
these findings support further investigation of urine-based multivariate
signatures, but differences in sample preparation, comparator groups
and validation design preclude direct performance comparisons. They
provide complementary biological and analytical context rather than
direct validation of the identity, native urinary origin or PC specificity
of the volatile and thermal-decomposition products assigned by THz
spectroscopy.
}

Given the complexity and heterogeneity of the urinary volatilome, which encompasses a wide array of chemical classes, it is evident that further research, particularly including machine learning approaches, will be essential to advance the identification of robust biomarkers for prostate pathology \cite{ref43_ProstateCancerScreenin}.

\begin{table*}[!t]
\centering
\begingroup
\fontsize{5.5}{5.5}\selectfont
\renewcommand{\arraystretch}{0.80}
\caption{Weak and strong potential prostate cancer markers in the BPH/PC dendrogram-derived molecule map. Strong potential markers are prioritised candidates for validation and should not yet be interpreted as diagnostic biomarkers.}\label{tab:weak-strong-pc-markers}
\begin{tabular}{@{}L{0.14\textwidth}L{0.16\textwidth}L{0.65\textwidth}@{}}
\toprule
Molecule & Candidate class & Reasoning \\
\midrule
Glycolaldehyde & Strong potential marker & Glycolaldehyde is one of the most biologically compelling prostate-cancer-associated candidates in the present dataset. Its detection in the prostate cancer group is supported by THz urine-vapour evidence \cite{ref26_ApplicatOfHigh} and by independent evidence that glycolaldehyde can occur in human urine in a cancer-risk clinical setting \cite{lin2021_quantSchema}. Its chemistry is also consistent with carbonyl stress, glycation chemistry and altered oxidative metabolism \cite{ref38_GlycolalAReactive}. Therefore, glycolaldehyde should be prioritised for targeted validation, while avoiding the stronger claim that it is prostate-cancer-specific at this stage. \\
Butyronitrile & Strong potential marker & Butyronitrile is a high-priority candidate because it was detected in the prostate cancer molecular fingerprint \cite{ref26_ApplicatOfHigh} and is not readily explained as a common urinary metabolite. Its nitrile chemistry may reflect disease-associated thermal decomposition products or altered nitrogen-containing urinary chemistry, an interpretation supported by urinary metabolomics evidence for nitrogen-derived volatile products \cite{ref37_UrinaryMetaboliBiomarke}. \\
Cyanovinylacetylene & Strong potential marker & Cyanovinylacetylene is a high-priority but high-caution candidate because it represents an unusual prostate-cancer-associated assignment in the THz fingerprint \cite{ref26_ApplicatOfHigh} with limited ordinary urinary precedent. Its value lies in its potential specificity within the THz fingerprint, but its chemical identity must be confirmed using multiple rotational transitions, isomer checks and standard-based validation. \\
Pentannitrile & Strong potential marker & Pentannitrile is a strong candidate because it belongs to the nitrile family and appears as part of the prostate-cancer-associated volatile pattern \cite{ref26_ApplicatOfHigh}. Since nitriles can plausibly arise from nitrogen-rich urinary substrates during thermal decomposition \cite{ref37_UrinaryMetaboliBiomarke}, pentannitrile may reflect a disease-associated chemical fingerprint rather than a conventional native metabolite. \\
Methyl isocyanate & Strong potential marker, high caution & Methyl isocyanate is retained as a strong but high-caution candidate because it is chemically unusual and potentially informative within the prostate cancer THz fingerprint \cite{ref26_ApplicatOfHigh}. However, its reactivity and possible thermal or environmental origins require particularly strict confirmation, including blank controls, standard spectra and assessment of sample-processing artifacts. \\
Malondialdehyde & Weak potential marker & Malondialdehyde supports an oxidative-stress interpretation of the prostate cancer urine-derived fingerprint, but it is not specific to prostate cancer. It is a well-known lipid-peroxidation product \cite{ref40_SerumMalondiaMda,ref53_UrinaryMalondiaMda} and can be influenced by inflammation, benign disease and systemic oxidative stress \cite{ref54_UrinaryInflammaAnd}. It should therefore be interpreted as supportive biological context rather than a standalone cancer marker. \\
Hydroxyacetone & Weak potential marker & Hydroxyacetone is consistent with a broader carbonyl/ketol metabolic signature and has independent support as a urine-associated carbonyl compound \cite{lin2021_quantSchema}. However, broader urinary metabolomics studies show that such metabolites can arise in complex urinary metabolic backgrounds \cite{ref19_TheUntargetUrine}, so its presence alone is insufficient to support prostate cancer specificity. \\
Dihydroxyacetone & Weak potential marker & Dihydroxyacetone is biologically plausible within carbohydrate and carbonyl metabolism. Prior urine-metabolite literature supports the broad relevance of carbohydrate-related urinary metabolites \cite{ref15_TheHumanUrine,ref46_NonInvasiveUrinary,ref47_1hNmrBased}, and carbonyl profiling supports its chemical context in urine-associated pathways \cite{lin2021_quantSchema}. Its relevance is therefore best framed as part of a broader metabolic disturbance rather than as an individual prostate cancer marker. \\
Benzaldehyde & Weak potential marker & Benzaldehyde has prior support in urinary and headspace volatilome studies \cite{ref19_TheUntargetUrine,ref48_UrinaryVolatileOrganic}, which makes its detection chemically plausible, and it has also been reported in THz urine-vapour work \cite{ref26_ApplicatOfHigh}. However, this prior support also limits its specificity, so it should not be interpreted as a standalone cancer marker without quantitative separation from benign disease. \\
Formic acid & Weak potential marker & Formic acid/formate is a common urinary and metabolic analyte \cite{ref15_TheHumanUrine,ref46_NonInvasiveUrinary} and can also reflect exposure-related chemistry, including formaldehyde or methanol-related pathways \cite{ref50_DeterminOfFormalde,ref52_UrinaryMethanolAnd}. Its detection may contribute to the overall chemical profile, but it is too nonspecific to be considered a strong prostate cancer marker by presence alone. \\
Glycine & Weak potential marker & Glycine is a common urinary amino acid with established roles in general metabolism \cite{ref15_TheHumanUrine,ref46_NonInvasiveUrinary}. Although NMR-based metabolomics supports its relevance to urinary metabolic profiling \cite{ref47_1hNmrBased}, its broad physiological occurrence makes it weak as an individual discriminatory marker. \\
Glycine amide & Weak potential marker & Glycine amide may reflect nitrogen-containing urinary chemistry and is compatible with the broader urinary metabolic context \cite{ref15_TheHumanUrine}. Its assignment in the present THz urine-vapour context provides exploratory support \cite{ref26_ApplicatOfHigh}, but because the interpretation depends strongly on assignment confidence and is not independently specific to prostate cancer, it is best treated as supportive rather than primary. \\
Methylphenyl ether & Weak potential marker & Methylphenyl ether is chemically plausible in a urine-derived volatile profile, because urinary and volatile metabolomics can include aromatic or exposure-influenced compounds \cite{ref15_TheHumanUrine,ref19_TheUntargetUrine}. Its THz urine-vapour assignment provides exploratory support \cite{ref26_ApplicatOfHigh}, but aromatic ether signals can be influenced by exogenous sources, environmental exposure or sample processing. \\
Phenol & Weak potential marker & Phenol is a known urinary and volatile compound \cite{ref15_TheHumanUrine,ref48_UrinaryVolatileOrganic} and has appeared in broader urinary metabolomics contexts \cite{ref19_TheUntargetUrine}. Because phenol may reflect diet, microbiome metabolism, medications, environmental exposure or disease-related metabolic change, it is useful as part of the overall volatilome pattern but weak as an individual prostate cancer marker. \\
Propanal & Weak potential marker & Propanal is a nonspecific volatile aldehyde that can arise from lipid oxidation, storage effects or general oxidative chemistry. Its relevance is supported by urinary metabolomics context \cite{ref19_TheUntargetUrine} and by literature on released volatile odorants \cite{ref49_MajorOdorantsReleased}, but it requires quantitative confirmation before any disease-specific interpretation is made. \\
\bottomrule
\end{tabular}
\endgroup
\end{table*}

\begin{table*}[!t]
\centering
\begingroup
\fontsize{5.5}{5.7}\selectfont
\renewcommand{\arraystretch}{0.82}
\captionof{table}{Comparison of selected prostate-cancer diagnostic techniques.}\label{tab:diagnostic-techniques}
\begin{tabular}{@{}>{\raggedright\arraybackslash}p{0.18\textwidth}>{\raggedright\arraybackslash}p{0.24\textwidth}>{\raggedright\arraybackslash}p{0.24\textwidth}>{\raggedright\arraybackslash}p{0.25\textwidth}@{}}
\toprule
Diagnostic Technique & Principle & Advantages & Limitations / References \\
\midrule
Prostate-Specific Antigen (PSA) Test & Serum total PSA \cite{ref5_BiologyOfProstate}. & Simple, inexpensive, and widely available. & Low specificity; benign conditions can elevate PSA \cite{ref6_DifferenDiagnosiOf}. \\
Prostate Health Index (PHI) & Composite of total PSA, free PSA, and [-2]proPSA \cite{ref7_DiagnostAbilityOf}. & Better PC-versus-benign discrimination than total PSA. & Requires blood sampling; inflammation can still produce false positives. \\
PCA3 (Prostate Cancer Antigen 3) Test & Urinary PCA3 non-coding RNA after digital rectal examination \cite{ref8_MoleculaPca3Diagnost,ref9_Pca3Tmprss2Erg}. & Non-invasive and more PC-specific than PSA. & Limited sensitivity; unsuitable for monitoring progression or aggressiveness. \\
Urinary microRNA Profiling & Cancer-related urinary microRNA expression \cite{ref10_Mir888Is,ref11_PotentiaUrinaryMirna,ref12_TheUtilityOf,ref13_DiagnostAndPrognost,ref14_AssessmeOfMir}. & Reflects tumor-associated molecular activity. & Complex RNA extraction; requires validation across populations. \\
Metabolomics (THz Urine Spectroscopy) & THz analysis of urine thermal-decomposition products \cite{ref19_TheUntargetUrine,ref20_NovelMetaboliSignatur,ref21_UntargetUrineMetaboli,ref22_The2017Terahert,ref24_ApplicatOfMicrowav,ref25_SensingNitrilesWith,ref26_ApplicatOfHigh}. & Non-invasive and label-free; detects biochemical differences between PC and BPH. & Developmental technology; requires standardization, spectral databases, and larger cohorts. \\
\bottomrule
\end{tabular}
\endgroup
\end{table*}

\section{Implementation Strategies and Technical Challenges}

Based on the findings of this pilot study, it is important to consider the practical steps required to translate high-resolution THz spectroscopy from a laboratory proof of concept to a clinically deployable diagnostic technology.

\begin{enumerate}
\item \textbf{Standardization of sample preparation;} Thermal decomposition and volatilization of urine metabolites must be performed under reproducible conditions of temperature, pressure, and dehydration rate. Small variations can alter the volatile fraction and affect the spectral reproducibility. Automated sample handling and controlled heating could support a standardized higher-throughput operation, but this workflow requires direct validation for THz urine spectroscopy.

\item \textbf{Sensitivity and selectivity improvements.} Many volatile products occur at trace concentrations. Enhancing source stability and detector sensitivity. For example, through heterodyne detection, cryogenic receivers or photonic integrated circuits can improve signal-to-noise performance \cite{ref22_The2017Terahert,ref23_TeraMirRadiatio,ref27_HighResolutiTerahert,ref28_OnThePossibil}. Chemometric approaches may also support compound identification in complex spectral mixtures, subject to independent validation.

\item \textbf{Portable and cost-effective instrumentation;} Current backward-wave-oscillator setups are bulky and require technical expertise. Advances in solid-state sources, including quantum-cascade lasers, superlattice multipliers, and photomixing, may enable compact, frequency-agile spectrometers \cite{ref30_GiantControllGigahert,ref31_AnalyticExpressiFor,ref32_HarmonicGeneratiIn,ref33_TerahertGeneratiBy}. Integration in photonic chips could support suitable devices for hospital laboratories, although the operating frequency range would need to be adapted \cite{ref22_The2017Terahert,ref23_TeraMirRadiatio}.

\item \textbf{Robust data analysis and clinical validation.} Translation into medical diagnostics requires adequately powered datasets and pre-specified analysis. The chemometric and machine-learning approaches used in GC--MS and NMR metabolomics may be adapted for THz spectral fingerprints \cite{ref37_UrinaryMetaboliBiomarke,ref42_IdentifiOfA}. {A recent methodological comparator is the serum-metabolomics reanalysis by Wang et al., which used the public MetaboLights dataset MTBLS6039
and reported nested cross-validation and SHAP-based model interpretation
\cite{Wang2026SerumSHAP}. Its relevance here is methodological rather
than direct evidence for urinary THz assignments. Future THz classifier
development should keep data-dependent preprocessing, feature selection
and model tuning within the training folds and distinguish internal
validation from independent prospective evaluation; SHAP-based feature
importance should not be interpreted as proof of biochemical mechanism
or molecular identity.} Open spectral databases and multicenter validation studies will be important for reproducibility and regulatory evaluation.

\item \textbf{Regulatory and interoperability frameworks;} Clinical translation will require the identification of applicable requirements in quality-management, electrical-safety, data-integrity, and traceability with regulatory specialists. Collaboration among physicists, clinicians, statisticians, and regulators can help define validation protocols and determine whether THz metabolomics can mature into a diagnostic platform that complements established biochemical tests.
\end{enumerate}

\section{Conclusion}
{High-resolution THz frequency-sweep spectroscopy provides a sensitive approach for investigating urine-derived volatile and thermal-decomposition products.

This review integrates the existing THz urine-spectroscopy literature with statistical reanalysis of the 24-PC/14-BPH clinical dataset and comparative interpretation of the associated molecular assignments. These observations remain exploratory and should not be interpreted as validated diagnostic biomarkers yet; confirmation in independent cohorts is required.}
\section*{Author Contributions}
Conceptualization: V. A. Atduev and A. V. Maslennikova;  Methodology: V. A. Atduev, A. V. Maslennikova and A. V. Maslennikova, E.G. Domracheva and V. L. Vaks; Investigation: V. A. Atduev, E. G. Domracheva , M. B. Chernyaeva , V. A. Anfertev, K. A. Atduev, A. V. Maslennikova and V. L. Vaks; Validation: A. V. Maslennikova , V. L. Vaks , Y. Tawalbeh and M. F. Pereira; Formal analysis: A.P. Bavrina, Y. Tawalbeh  and M. F. Pereira;  Supervision: A. V. Maslennikova , V. L. Vaks and M. F. Pereira; Funding acquisition: V. L. Vaks;  Visualization: Y. Tawalbeh; Project administration: V. L. Vaks and A. V. Maslennikova;  All authors contributed to writing the original draft, reviewing and editing.

\section*{Funding}

The studies of biological-fluid samples aimed at identifying disease markers were carried out under State Assignment FFUF-2024-0024. 

\section*{Institutional Review Board Statement}

The study was conducted according to the Declaration of Helsinki guidelines and was approved by the Institutional Review Board of the Nizhny Novgorod Regional Oncology Hospital (protocol 5, 18 May 2015).

\section*{Informed Consent Statement}

Written informed consent was obtained from all participants involved in the study.

\section*{Data Availability Statement}

The data that support the findings of this study are available on request from the corresponding author. The data are not publicly available due to privacy or ethical restrictions.

\section*{Conflicts of Interest}

The authors declare no conflicts of interest.

\clearpage
\onecolumn
\section*{Supporting Information}
\captionsetup[table]{labelfont=bf,labelsep=colon}
\renewcommand{\thetable}{S\arabic{table}}
\setcounter{table}{0}
\setlength{\tabcolsep}{2pt}
\setlength{\LTpre}{0.5em}
\setlength{\LTpost}{0.5em}

\begin{scriptsize}
\begin{longtable}{>{\raggedright\arraybackslash}p{0.16\linewidth}>{\raggedright\arraybackslash}p{0.25\linewidth}>{\raggedright\arraybackslash}p{0.15\linewidth}>{\raggedright\arraybackslash}p{0.14\linewidth}>{\raggedright\arraybackslash}p{0.14\linewidth}>{\raggedright\arraybackslash}p{0.06\linewidth}}
\caption{Substances present in both PC and BPH samples}\label{tab:s1}\\
\toprule
Substance & Formula & Experimental frequency (MHz) & Catalogue frequency (MHz) & Intensity lg(nm\textasciicircum{}2*MHz) & Line count \\
\midrule
\endfirsthead
\caption[]{Substances present in both PC and BPH samples (continued)}\\
\toprule
Substance & Formula & Experimental frequency (MHz) & Catalogue frequency (MHz) & Intensity lg(nm\textasciicircum{}2*MHz) & Line count \\
\midrule
\endhead
\midrule
\multicolumn{6}{r}{\emph{Continued on next page}}\\
\endfoot
\bottomrule
\endlastfoot
Isocyanic acid & HNCO, v = 0 & 131394.4 & 131394.2302 & -3.3018 & 20.0 \\
 &  & 131640.9 & 131640.7705 & -5.3043 &  \\
 &  & 131733.9 & 131733.5888 & -4.5002 &  \\
 &  & 131799.4 & 131799.2971 & -3.9329 &  \\
 &  & 131886.0 & 131885.7341 & -3.2188 &  \\
 &  & 132115.4 & 132115.2675 & -4.2568 &  \\
 &  & 132356.9 & 132356.7014 & -3.2957 &  \\
 &  & 134481.8 & 134481.5722 & -3.7314 &  \\
 &  & 138909.9 & 138909.7510 & -4.2115 &  \\
 &  & 145068.9 & 145068.5389 & -3.6246 &  \\
 &  & 149257.5 & 149257.485 & -3.0223 &  \\
 &  & 153292.1 & 153291.9354 & -3.1076 &  \\
 &  & 153764.9 & 153764.6059 & -3.7052 &  \\
 &  & 153865.1 & 153865.0858 & -3.0279 &  \\
 &  & 154414.8 & 154414.7653 & -3.1015 &  \\
 &  & 158934.4 & 158934.8846 & -3.5512 &  \\
 &  & 164144.0 & 164143.9173 & -4.1234 &  \\
 &  & 166940.4 & 166940.5189 & -5.2509 &  \\
 &  & 169954.4 & 169954.1768 & -2.9313 &  \\
 &  & 170578.0 & 170578.0448 & -2.8594 &  \\
Acetic acid & CH3C(O)OH, vt=0 & 122199.5 & 122199.4324 & -5.0494 & 25 \\
 &  & 131292.6 & 131292.385 & -5.2909 &  \\
 &  & 131675.5 & 131675.3307 & -5.5715 &  \\
 &  & 133130.5 & 133130.4321 & -5.0853 &  \\
 &  & 133191.1 & 133191.0185 & -5.3463 &  \\
 &  & 134258.9 & 134258.92 & -5.1982 &  \\
 &  & 134315.9 & 134315.7594 & -5.2294 &  \\
 &  & 137193.8 & 137193.8686 & -5.3539 &  \\
 &  & 138251.3 & 138251.1805 & -5.3605 &  \\
 &  & 141876.5 & 141876.7391 & -5.5459 &  \\
 &  & 143461.7 & 143461.5241 & -4.7948 &  \\
 &  & 144791.8 & 144791.7911 & -5.0981 &  \\
 &  & 148326.4 & 148326.4548 & -5.0999 &  \\
 &  & 149952.6 & 149952.3063 & -4.9916 &  \\
 &  & 150393.9 & 150394.0813 & -5.2044 &  \\
 &  & 151549.9 & 151549.913 & -5.1992 &  \\
 &  & 159304.9 & 159304.6114 & -5.3042 &  \\
 &  & 160660.4 & 160660.3784 & -4.9122 &  \\
 &  & 164293.1 & 164292.7405 & -4.8064 &  \\
 &  & 164763.1 & 164762.926 & -4.7312 &  \\
 &  & 164800.9 & 164800.4866 & -4.7036 &  \\
 &  & 164927.5 & 164927.5635 & -5.9794 &  \\
 &  & 166295.0 & 166295.1272 & -4.9904 &  \\
 &  & 174711.3 & 174711.5366 & -4.9614 &  \\
 &  & 174841.5 & 174841.2244 & -5.0556 &  \\
Propanediol & gG'a-CH3CHOHCH2OH, v = 0 & 131606.5 & 131606.4496 & -5.3653 &  12 \\
 &  & 131612.4 & 131612.431 & -4.9122 &  \\
 &  & 131846.4 & 131846.2918 & -3.9839 &  \\
 &  & 153653.5 & 153653.1389 & -5.9165 &  \\
 &  & 154227.9 & 154227.8494 & -5.3495 &  \\
 &  & 156905.6 & 156905.2392 & -4.4959 &  \\
 &  & 158151.7 & 158151.7621 & -5.473 &  \\
 &  & 166584.6 & 166584.208 & -5.7984 &  \\
 &  & 167298.2 & 167298.3983 & -4.3183 &  \\
 &  & 168722.2 & 168722.082 & -4.3076 &  \\
 &  & 172169.4 & 172169.4536 & -4.9923 &  \\
 &  & 174966.9 & 174966.8944 & -4.6122 &  \\
Acetaldehyde & CH3CHO, vt = 0, 1, 2 & 118592.2 & 118591.9643 & -4.1274 & 32 \\
 &  & 130892.3 & 130892.749 & -3.9785 &  \\
 &  & 134075.1 & 134075.3798 & -4.2428 &  \\
 &  & 134694.6 & 134694.4463 & -3.9924 &  \\
 &  & 134895.6 & 134895.606 & -3.9997 &  \\
 &  & 134908.7 & 134908.6558 & -4.1647 &  \\
 &  & 134963.7 & 134963.2399 & -4.0581 &  \\
 &  & 134973.3 & 134973.0698 & -4.058 &  \\
 &  & 134987.6 & 134987.2121 & -4.0579 &  \\
 &  & 134996.1 & 134996.0633 & -4.0578 &  \\
 &  & 135083.7 & 135083.2542 & -4.5564 &  \\
 &  & 135685.9 & 135685.4741 & -3.9861 &  \\
 &  & 135804.3 & 135804.3514 & -4.2807 &  \\
 &  & 137045.5 & 137045.3105 & -4.2445 &  \\
 &  & 138319.6 & 138319.628 & -3.9325 &  \\
 &  & 149504.9 & 149505.1281 & -3.8127 &  \\
 &  & 149507.6 & 149507.4617 & -3.8128 &  \\
 &  & 149687.1 & 149686.9142 & -4.1056 &  \\
 &  & 150084.3 & 150084.3242 & -5.0578 &  \\
 &  & 152936.2 & 152936.5451 & -4.081 &  \\
 &  & 153969.7 & 153969.8346 & -4.1147 &  \\
 &  & 154146.0 & 154145.5647 & -4.5669 &  \\
 &  & 154274.9 & 154274.6864 & -3.8716 &  \\
 &  & 157666.2 & 157666.5805 & -6.0035 &  \\
 &  & 171265.9 & 171265.4706 & -3.6465 &  \\
 &  & 171297.3 & 171296.9846 & -3.6465 &  \\
 &  & 173519.3 & 173519.1037 & -3.7823 &  \\
 &  & 173534.0 & 173534.2359 & -3.7823 &  \\
 &  & 173638.0 & 173638.0662 & -3.7147 &  \\
 &  & 173664.1 & 173663.5315 & -3.7146 &  \\
 &  & 173682.7 & 173682.4083 & -3.7143 &  \\
 &  & 174706.7 & 174706.7017 & -3.9613 &  \\
Acetonitrile & CH3CN, v = 0 & 128758.2 & 128757.0304 & -3.0163 & 25 \\
 &  & 128769.6 & 128769.4365 & -2.9134 &  \\
 &  & 128777.1 & 128776.8821 & -2.8543 &  \\
 &  & 128779.3 & 128779.3643 & -2.8349 &  \\
 &  & 129043.3 & 129043.162 & -3.5976 &  \\
 &  & 147149.1 & 147149.0687 & -2.8297 &  \\
 &  & 147163.4 & 147163.2445 & -2.7401 &  \\
 &  & 147171.7 & 147171.7524 & -2.6878 &  \\
 &  & 147174.7 & 147174.5887 & -2.6706 &  \\
 &  & 147610.8 & 147611.034 & -3.4325 &  \\
 &  & 147619.9 & 147619.916 & -3.4633 &  \\
 &  & 147760.6 & 147760.654 & -3.4295 &  \\
 &  & 165454.0 & 165454.3701 & -3.156 &  \\
 &  & 165489.3 & 165489.3912 & -2.9472 &  \\
 &  & 165518.1 & 165518.0642 & -2.7893 &  \\
 &  & 165556.2 & 165556.3223 & -2.5914 &  \\
 &  & 165565.7 & 165565.8918 & -2.5438 &  \\
 &  & 165569.2 & 165569.082 & -2.528 &  \\
 &  & 165907.8 & 165908.1411 & -3.2871 &  \\
 &  & 166013.5 & 166013.4004 & -3.4921 &  \\
 &  & 166036.0 & 166035.8883 & -3.3919 &  \\
 &  & 166051.5 & 166051.5011 & -3.325 &  \\
 &  & 166059.1 & 166058.8124 & -3.2899 &  \\
 &  & 166071.1 & 166071.2258 & -3.3147 &  \\
 &  & 166228.3 & 166228.3145 & -3.2855 &  \\
Methylbutironitrile & C2H5CHCNCH3, v = 0 & 128017.1 & 128016.8668 & -5.6251 & 11 \\
 &  & 130974.9 & 130975.286 & -5.3136 &  \\
 &  & 131952.6 & 131952.4245 & -5.1574 &  \\
 &  & 133795.9 & 133795.9682 & -3.7732 &  \\
 &  & 135531.8 & 135531.9312 & -4.1832 &  \\
 &  & 153414.0 & 153414.274 & -4.298 &  \\
 &  & 153483.7 & 153483.672 & -4.4746 &  \\
 &  & 162210.5 & 162210.2023 & -5.0059 &  \\
 &  & 163132.4 & 163131.9561 & -5.564 &  \\
 &  & 166626.7 & 166626.7683 & -5.9879 &  \\
 &  & 173054.5 & 173054.8521 & -5.9271 &  \\
Pentadiene nitrile & E-CH2(CH)3CN, v = 0 & 124284.5 & 124284.5804 & -5.1411 & 18 \\
 &  & 128605.3 & 128605.4944 & -4.3189 &  \\
 &  & 130719.7 & 130719.7656 & -4.3232 &  \\
 &  & 130988.0 & 130987.7565 & -5.3094 &  \\
 &  & 131730.8 & 131730.3439 & -5.7456 &  \\
 &  & 135872.7 & 135872.8822 & -5.8211 &  \\
 &  & 136156.9 & 136157.1538 & -5.2123 &  \\
 &  & 136583.7 & 136583.7229 & -4.6669 &  \\
 &  & 138531.3 & 138531.2773 & -4.2545 &  \\
 &  & 147051.1 & 147050.2923 & -5.6007 &  \\
 &  & 150006.8 & 150007.2046 & -5.2165 &  \\
 &  & 153388.9 & 153388.8379 & -3.7691 &  \\
 &  & 156012.9 & 156013.0571 & -5.8105 &  \\
 &  & 165069.0 & 165069.0687 & -4.4538 &  \\
 &  & 168349.3 & 168349.8118 & -3.4189 &  \\
 &  & 173343.7 & 173343.7229 & -5.507 &  \\
 &  & 173500.0 & 173499.7023 & -3.458 &  \\
 &  & 173671.6 & 173671.7013 & -4.9633 &  \\
Ethynyl benzonitrile & 1-CN-3-CCH-C6H4, v = 0 & 140983.5 & 140983.254 & -4.2355 & 7 \\
 &  & 141070.9 & 141070.837 & -5.4492 &  \\
 &  & 144221.4 & 144221.21 & -4.7897 &  \\
 &  & 154142.5 & 154142.2844 & -5.2937 &  \\
 &  & 155033.5 & 155033.49 & -4.2629 &  \\
 &  & 172174.9 & 172175.309 & -4.019 &  \\
 &  & 173236.1 & 173235.8408 & -5.3961 &  \\
Methyl mercaptan & CH3SH, v = 0, 1, 2 & 125131.3 & 125130.8629 & -4.4675 & 23 \\
 &  & 125968.4 & 125968.4654 & -4.9393 &  \\
 &  & 126195.3 & 126194.6129 & -4.4601 &  \\
 &  & 126403.9 & 126403.8344 & -4.4357 &  \\
 &  & 126406.0 & 126405.6761 & -4.4339 &  \\
 &  & 127745.8 & 127745.8642 & -4.4499 &  \\
 &  & 134686.3 & 134686.3683 & -5.514 &  \\
 &  & 135880.8 & 135880.4684 & -4.621 &  \\
 &  & 149777.8 & 149778.3263 & -4.9595 &  \\
 &  & 150147.1 & 150146.933 & -4.2341 &  \\
 &  & 150205.3 & 150205.3581 & -4.6618 &  \\
 &  & 151136.0 & 151135.7238 & -4.6871 &  \\
 &  & 151303.2 & 151303.1941 & -4.2276 &  \\
 &  & 151311.0 & 151311.0125 & -4.5614 &  \\
 &  & 151654.2 & 151654.2179 & -4.2079 &  \\
 &  & 151660.0 & 151660.0468 & -4.2062 &  \\
 &  & 151721.1 & 151720.7511 & -4.5616 &  \\
 &  & 151733.9 & 151733.6476 & -4.3869 &  \\
 &  & 151738.1 & 151737.4068 & -4.3887 &  \\
 &  & 151760.1 & 151759.9646 & -4.2828 &  \\
 &  & 151794.9 & 151794.8504 & -4.284 &  \\
 &  & 152129.5 & 152129.0284 & -4.2253 &  \\
 &  & 166578.9 & 166578.4447 & -4.6503 &  \\
Alanine & CH3CHNH2COOH, v = 0 & 136269.0 & 136268.5992 & -5.1532 & 5 \\
 &  & 144975.3 & 144975.4725 & -5.2424 &  \\
 &  & 154119.1 & 154118.9248 & -5.3674 &  \\
 &  & 157974.7 & 157974.786 & -5.7456 &  \\
 &  & 163322.9 & 163322.733 & -3.3239 &  \\
Urea & H2NC(O)NH2, v2\textbackslash{} \& v3 & 127063.4 & 127063.5785 & -4.7559 & 12 \\
 &  & 132196.4 & 132195.8689 & -4.2114 &  \\
 &  & 132713.2 & 132712.5176 & -5.2501 &  \\
 &  & 132850.0 & 132849.8463 & -4.4258 &  \\
 &  & 134259.0 & 134259.0507 & -4.8767 &  \\
 &  & 135013.5 & 135014.1423 & -4.1356 &  \\
 &  & 153956.4 & 153956.5079 & -3.5987 &  \\
 &  & 157711.7 & 157711.6818 & -5.4001 &  \\
 &  & 163380.2 & 163379.9247 & -3.6655 &  \\
 &  & 163759.2 & 163759.5911 & -4.1441 &  \\
 &  & 164062.8 & 164062.9567 & -4.1153 &  \\
 &  & 164996.8 & 164996.485 & -4.4363 &  \\
Ethylene glycol & gGg'-(CH2OH)2, v = 0 & 120289.9 & 120289.7710 & -5.2756 & 8 \\
 &  & 131303.1 & 131303.289 & -4.7072 &  \\
 &  & 131868.6 & 131868.4654 & -5.4247 &  \\
 &  & 135004.7 & 135005.56 & -4.5619 &  \\
 &  & 153166.3 & 153166.2125 & -5.1791 &  \\
 &  & 153819.5 & 153819.1141 & -6.2446 &  \\
 &  & 154142.5 & 154142.2026 & -5.1221 &  \\
 &  & 173813.4 & 173813.1032 & -4.8908 &  \\
Methyl carbamate & H2NCO2CH3, v = 0 & 124746.3 & 124746.1953 & -5.4106 & 11 \\
 &  & 132810.5 & 132810.9433 & -5.5071 &  \\
 &  & 133830.5 & 133830.8697 & -5.1441 &  \\
 &  & 136697.1 & 136696.9244 & -4.8323 &  \\
 &  & 140796.3 & 140796.5902 & -5.1771 &  \\
 &  & 141862.9 & 141862.8625 & -5.1465 &  \\
 &  & 150112.1 & 150112.1892 & -5.2834 &  \\
 &  & 152891.2 & 152891.2741 & -5.288 &  \\
 &  & 162154.5 & 162154.4922 & -5.2726 &  \\
 &  & 165956.4 & 165956.4103 & -5.2488 &  \\
 &  & 166211.4 & 166211.42 & -5.729 &  \\
Cyanoethynylbenzole & C6H5C3N, v = 0 & 132501.8 & 132501.4714 & -3.9968 & 7 \\
 &  & 134524.4 & 134524.0023 & -5.6197 &  \\
 &  & 136480.5 & 136480.6169 & -4.485 &  \\
 &  & 152411.4 & 152410.939 & -3.6493 &  \\
 &  & 152635.2 & 152634.796 & -3.6272 &  \\
 &  & 155027.2 & 155026.942 & -3.6978 &  \\
 &  & 158853.6 & 158853.5624 & -5.1485 &  \\
Carbonyl sulphide & OCS, v = 0 & 121625.1 & 121624.6380 & -3.3522 & 7 \\
 &  & 133786.1 & 133785.9000 & -3.2369 &  \\
 &  & 134089.5 & 134088.701 & -4.3359 &  \\
 &  & 146124.6 & 146124.4576 & -4.2326 &  \\
 &  & 146277.4 & 146277.1016 & -4.2318 &  \\
 &  & 158107.7 & 158107.36 & -3.0396 &  \\
 &  & 170267.6 & 170267.4940 & -2.9544 &  \\
Sulphur dioxide & SO2, v = 0 & 118577.3 & 118577.43 & -4.212 & 36 \\
 &  & 123058.0 & 123057.69 & -4.581 &  \\
 &  & 124865.3 & 124864.74 & -4.332 &  \\
 &  & 129105.7 & 129105.83 & -3.9128 &  \\
 &  & 129515.1 & 129514.81 & -3.5248 &  \\
 &  & 130859.8 & 130859.4 & -4.3862 &  \\
 &  & 131014.9 & 131014.86 & -3.6063 &  \\
 &  & 131275.3 & 131274.93 & -4.2324 &  \\
 &  & 131530.7 & 131530.479 & -4.5218 &  \\
 &  & 132745.2 & 132744.86 & -3.3945 &  \\
 &  & 133003.8 & 133003.6131 & -4.6728 &  \\
 &  & 134005.0 & 134004.86 & -3.6056 &  \\
 &  & 134943.7 & 134943.3 & -4.2463 &  \\
 &  & 135696.1 & 135696.02 & -3.8143 &  \\
 &  & 135963.2 & 135963.0 & -4.4389 &  \\
 &  & 136674.8 & 136675.4059 & -4.8907 &  \\
 &  & 139474.4 & 139474.54 & -4.3251 &  \\
 &  & 140306.6 & 140306.17 & -3.7171 &  \\
 &  & 143057.2 & 143057.11 & -3.3274 &  \\
 &  & 146550.2 & 146550.08 & -4.2474 &  \\
 &  & 146605.5 & 146605.52 & -3.891 &  \\
 &  & 147590.0 & 147589.9176 & -4.323 &  \\
 &  & 150381.1 & 150381.1 & -4.1363 &  \\
 &  & 151379.0 & 151378.63 & -4.2703 &  \\
 &  & 157135.3 & 157135.2524 & -4.3148 &  \\
 &  & 159448.0 & 159447.96 & -4.1989 &  \\
 &  & 159887.8 & 159887.2982 & -4.4913 &  \\
 &  & 160343.2 & 160342.99 & -3.2564 &  \\
 &  & 160827.8 & 160827.88 & -3.4028 &  \\
 &  & 163119.7 & 163119.3789 & -3.8035 &  \\
 &  & 163605.8 & 163605.5328 & -3.4644 &  \\
 &  & 163924.9 & 163924.7288 & -4.4005 &  \\
 &  & 165144.9 & 165144.6507 & -3.7597 &  \\
 &  & 165225.6 & 165225.4511 & -3.5083 &  \\
 &  & 165266.4 & 165266.215 & -4.6419 &  \\
 &  & 170754.3 & 170754.5483 & -4.0546 &  \\
\end{longtable}
\end{scriptsize}

\begin{scriptsize}
\begin{longtable}{>{\raggedright\arraybackslash}p{0.16\linewidth}>{\raggedright\arraybackslash}p{0.25\linewidth}>{\raggedright\arraybackslash}p{0.14\linewidth}>{\raggedright\arraybackslash}p{0.14\linewidth}>{\raggedright\arraybackslash}p{0.16\linewidth}>{\raggedright\arraybackslash}p{0.06\linewidth}}
\caption{Substances present in PC samples but absent in BPH samples}\label{tab:s2}\\
\toprule
Substance & Formula & Experimental frequency (MHz) & Catalogue frequency (MHz) & Intensity lg(nm\textasciicircum{}2*MHz) & Line count \\
\midrule
\endfirsthead
\caption[]{Substances present in PC samples but absent in BPH samples (continued)}\\
\toprule
Substance & Formula & Experimental frequency (MHz) & Catalogue frequency (MHz) & Intensity lg(nm\textasciicircum{}2*MHz) & Line count \\
\midrule
\endhead
\midrule
\multicolumn{6}{r}{\emph{Continued on next page}}\\
\endfoot
\bottomrule
\endlastfoot
Formic acid & t-HCOOH, v = 0 & 129671.9 & 129671.8163 & -3.8608 & 13 \\
 &  & 133767.3 & 133767.1873 & -3.8183 &  \\
 &  & 134686.3 & 134686.37 & -3.8963 &  \\
 &  & 134938.7 & 134938.3853 & -4.1392 &  \\
 &  & 135737.7 & 135737.7589 & -3.8749 &  \\
 &  & 151176.3 & 151176.2809 & -3.6818 &  \\
 &  & 155617.9 & 155617.8821 & -3.645 &  \\
 &  & 157054.2 & 157053.9807 & -3.6919 &  \\
 &  & 157214.7 & 157214.5633 & -6.1522 &  \\
 &  & 157526.8 & 157526.5212 & -3.7635 &  \\
 &  & 158720.7 & 158720.5303 & -3.6682 &  \\
 &  & 162598.5 & 162598.4803 & -3.6214 &  \\
 &  & 172635.7 & 172635.7787 & -3.5177 &  \\
Phenol & c-C6H5OH, v = 0 & 152589.1 & 152588.874 & -5.5305 & 3 \\
 &  & 152788.7 & 152788.828 & -5.0252 &  \\
 &  & 168879.5 & 168879.6808 & -5.8954 &  \\
Methylphenylether & c-C6H5OCH3, v = 0 & 131332.9 & 131332.6094 & -5.8198 & 3 \\
 &  & 132100.0 & 132100.3050 & -5.3989 &  \\
 &  & 170850.4 & 170850.3602 & -5.851 &  \\
Propanal & s-C2H5CHO, v = 0 & 132322.4 & 132322.8377 & -5.809 & 6 \\
 &  & 136243.7 & 136243.3931 & -4.9183 &  \\
 &  & 157962.2 & 157961.8463 & -4.8524 &  \\
 &  & 163315.4 & 163315.767 & -4.4344 &  \\
 &  & 168298.0 & 168297.913 & -4.8963 &  \\
 &  & 168904.1 & 168903.3942 & -4.3562 &  \\
Benzaldehyde & c-C6H5CHO, v = 0 & 133854.5 & 133854.1009 & -5.3594 & 5 \\
 &  & 138582.0 & 138582.0441 & -3.9696 &  \\
 &  & 139438.0 & 139438.1285 & -3.985 &  \\
 &  & 147058.8 & 147058.5841 & -5.3405 &  \\
 &  & 168493.3 & 168492.8956 & -5.6822 &  \\
Glycolaldehyde & CH2(OH)CHO, v = 0 & 122687.5 & 121686.6661 & -3.9287 & 93 \\
 &  & 123430.9 & 123430.7543 & -3.959 &  \\
 &  & 124036.3 & 124035.9593 & -4.5438 &  \\
 &  & 124043.7 & 124043.7677 & -3.8454 &  \\
 &  & 125780.9 & 125780.992 & -4.071 &  \\
 &  & 126236.2 & 126236.1983 & -3.8811 &  \\
 &  & 127294.1 & 127294.0386 & -4.1508 &  \\
 &  & 127708.3 & 127708.0323 & -3.9539 &  \\
 &  & 129902.8 & 129902.6816 & -4.0313 &  \\
 &  & 131008.4 & 131008.3007 & -3.9574 &  \\
 &  & 131746.4 & 131746.2388 & -3.8689 &  \\
 &  & 131963.6 & 131963.4451 & -3.8846 &  \\
 &  & 133282.1 & 133282.1215 & -3.8509 &  \\
 &  & 133411.0 & 133410.6579 & -4.1212 &  \\
 &  & 133454.1 & 133453.6955 & -3.9695 &  \\
 &  & 133485.5 & 133485.5473 & -4.1837 &  \\
 &  & 133735.6 & 133735.6004 & -4.182 &  \\
 &  & 133886.5 & 133886.1018 & -3.7478 &  \\
 &  & 134335.9 & 134335.9526 & -4.3978 &  \\
 &  & 134743.3 & 134743.3845 & -3.8961 &  \\
 &  & 134884.1 & 134883.6226 & -3.9135 &  \\
 &  & 134985.9 & 134985.8225 & -4.9752 &  \\
 &  & 135205.0 & 135204.9264 & -4.3265 &  \\
 &  & 135222.3 & 135222.1851 & -3.99 &  \\
 &  & 135300.3 & 135299.8979 & -3.9331 &  \\
 &  & 135767.4 & 135767.4934 & -3.9233 &  \\
 &  & 135833.5 & 135833.3572 & -3.8645 &  \\
 &  & 136487.5 & 136487.2961 & -4.0184 &  \\
 &  & 136504.6 & 136504.261 & -3.9827 &  \\
 &  & 136989.9 & 136989.8466 & -3.8713 &  \\
 &  & 137123.7 & 137123.5581 & -4.0148 &  \\
 &  & 137148.7 & 137148.8347 & -3.9644 &  \\
 &  & 137333.2 & 137332.9613 & -3.849 &  \\
 &  & 137390.4 & 137390.6949 & -4.0553 &  \\
 &  & 137677.1 & 137677.0979 & -3.9239 &  \\
 &  & 137684.0 & 137684.2168 & -4.0536 &  \\
 &  & 137842.9 & 137843.0313 & -3.9094 &  \\
 &  & 138037.5 & 138037.2383 & -4.1017 &  \\
 &  & 138547.9 & 138548.2835 & -4.1599 &  \\
 &  & 139158.4 & 139158.2323 & -3.8311 &  \\
 &  & 139727.4 & 139727.6569 & -3.8761 &  \\
 &  & 139818.2 & 139818.1149 & -3.8243 &  \\
 &  & 139946.0 & 139946.1264 & -3.8727 &  \\
 &  & 143765.9 & 143765.7936 & -3.6588 &  \\
 &  & 149616.7 & 149616.5437 & -4.6136 &  \\
 &  & 149751.5 & 149751.3904 & -4.1878 &  \\
 &  & 149966.9 & 149966.9587 & -3.9555 &  \\
 &  & 150034.6 & 150034.492 & -3.7907 &  \\
 &  & 150460.0 & 150459.5261 & -4.2156 &  \\
 &  & 151294.0 & 151294.2558 & -3.8294 &  \\
 &  & 151856.9 & 151856.9262 & -3.7793 &  \\
 &  & 151916.2 & 151916.4841 & -4.3568 &  \\
 &  & 152231.0 & 152231.0693 & -3.8402 &  \\
 &  & 152290.7 & 152290.694 & -3.5904 &  \\
 &  & 153598.0 & 153598.0148 & -3.5779 &  \\
 &  & 153667.0 & 153666.9684 & -3.5775 &  \\
 &  & 154393.5 & 154393.3571 & -3.8919 &  \\
 &  & 154777.9 & 154777.6023 & -3.7558 &  \\
 &  & 154828.7 & 154828.274 & -5.4922 &  \\
 &  & 154848.0 & 154847.7023 & -3.7312 &  \\
 &  & 155308.6 & 155308.5499 & -3.9155 &  \\
 &  & 156342.1 & 156342.0426 & -3.7862 &  \\
 &  & 156471.9 & 156471.562 & -3.7431 &  \\
 &  & 156664.2 & 156664.0975 & -3.7486 &  \\
 &  & 156775.7 & 156775.2942 & -3.7379 &  \\
 &  & 157966.3 & 157966.4126 & -4.1845 &  \\
 &  & 157992.4 & 157992.2897 & -3.7632 &  \\
 &  & 160493.8 & 160493.7897 & -3.8761 &  \\
 &  & 161835.0 & 161834.5507 & -3.6586 &  \\
 &  & 161952.0 & 161951.8083 & -3.7198 &  \\
 &  & 162734.6 & 162734.8683 & -3.8813 &  \\
 &  & 162879.4 & 162879.5684 & -3.9925 &  \\
 &  & 163074.6 & 163074.6577 & -3.7867 &  \\
 &  & 163349.7 & 163349.8341 & -3.9371 &  \\
 &  & 163580.2 & 163580.1028 & -3.5029 &  \\
 &  & 164047.5 & 164047.071 & -3.6458 &  \\
 &  & 164209.6 & 164209.6024 & -4.035 &  \\
 &  & 164215.5 & 164215.1767 & -3.7965 &  \\
 &  & 164432.8 & 164432.5534 & -3.7404 &  \\
 &  & 164446.1 & 164446.1844 & -4.201 &  \\
 &  & 165418.2 & 165418.0609 & -3.7119 &  \\
 &  & 166319.8 & 166320.2668 & -5.8997 &  \\
 &  & 167587.1 & 167587.2743 & -3.698 &  \\
 &  & 170570.4 & 170570.4856 & -3.8397 &  \\
 &  & 171751.3 & 171751.316 & -3.7609 &  \\
 &  & 172112.2 & 172112.0513 & -4.4467 &  \\
 &  & 172214.5 & 172214.4546 & -4.1913 &  \\
 &  & 172360.1 & 172360.7612 & -3.679 &  \\
 &  & 172791.9 & 172791.5846 & -3.9575 &  \\
 &  & 173507.6 & 173507.5245 & -3.5672 &  \\
 &  & 173953.7 & 173953.8331 & -3.8873 &  \\
 &  & 174114.1 & 174113.8879 & -3.7458 &  \\
 &  & 174130.1 & 174129.9 & -3.7544 &  \\
Malone dialdehyde & HOCHCHCHO, v = 0 & 125968.7 & 125968.568 & -4.2822 & 4 \\
 &  & 136687.6 & 136687.4573 & -4.6361 &  \\
 &  & 147286.2 & 147286.2491 & -5.4244 &  \\
 &  & 173071.4 & 173071.2432 & -4.5486 &  \\
Butyronitrile & i-C3H7CN, v = 0 & 132125.7 & 132125.8804 & -3.8966 & 8 \\
 &  & 133434.9 & 133434.8544 & -4.7453 &  \\
 &  & 134549.2 & 134549.4887 & -3.8619 &  \\
 &  & 145671.5 & 145671.478 & -4.8563 &  \\
 &  & 147476.1 & 147475.6457 & -5.9575 &  \\
 &  & 148781.7 & 148781.6479 & -3.4027 &  \\
 &  & 149971.0 & 149970.6108 & -5.0269 &  \\
 &  & 151243.1 & 151243.1824 & -3.4963 &  \\
Penannitrile & AG-n-C4H9CN, v = 0 & 128022.4 & 128022.2779 & -5.9304 & 8 \\
 &  & 133662.3 & 133662.8866 & -4.6063 &  \\
 &  & 136332.5 & 136332.5152 & -5.6038 &  \\
 &  & 140354.2 & 140354.0523 & -3.6819 &  \\
 &  & 145090.8 & 145090.7461 & -3.7777 &  \\
 &  & 150792.4 & 150792.325 & -4.1246 &  \\
 &  & 173854.9 & 173854.859 & -4.3949 &  \\
 &  & 174210.1 & 174210.2218 & -4.1086 &  \\
Methyl isocyanate & CH3NCO, v = 0 & 165968.9 & 165969.028 & -4.0874 & 3 \\
 &  & 173025.7 & 173025.646 & -4.2777 &  \\
 &  & 174161.2 & 174161.002 & -3.8161 &  \\
\end{longtable}
\end{scriptsize}
\newpage


\begin{table}[!tp]
\centering
\small
\caption{Related reports on high-resolution THz spectroscopy of urine-derived products: study scope, reported evidence and considerations for interpretation. Reports from the same research programme are not assumed to represent independent participant cohorts.}\label{tab:s3}
\footnotesize 
\label{tab:evidence-map}
\vspace{4pt}
\setlength{\tabcolsep}{4pt}
\renewcommand{\arraystretch}{1.15}
\begin{tabularx}{\textwidth}{@{}
  >{\raggedright\arraybackslash}p{0.15\textwidth}
  >{\raggedright\arraybackslash}p{0.18\textwidth}
  >{\raggedright\arraybackslash}X
  >{\raggedright\arraybackslash}X
  >{\raggedright\arraybackslash}p{0.16\textwidth}
@{}}
\toprule
\textbf{Study / report} &
\textbf{Cohort / scope} &
\textbf{Evidence reported} &
\textbf{Limitations / caveats} &
\textbf{Role in review} \\
\midrule
Vaks et al., 2022 \cite{ref25_SensingNitrilesWith}
& Cancer patients during chemotherapy
& Longitudinal clinical and spectroscopy observations reported in the article
& Different clinical question; not a PC-versus-BPH diagnostic cohort
& Technical and biological precursor \\
\addlinespace[4pt]
Vaks et al., 2024, \emph{Applied Sciences} \cite{ref26_ApplicatOfHigh}
& 12 PC versus 4 conditionally healthy men aged 25--34
& Patient-by-molecule table with 33 unique molecule rows
& Very small, age-mismatched control group; one blank cell; no external validation
& Foundational PC feature matrix \\
\addlinespace[4pt]
Vaks et al., 2024, IEEE proceedings \cite{ref44_Vaks2024IEEE}
& PC and BPH
& Conference report and group-level claims
& Full patient-level matrix not available in the material reviewed
& Intermediate PC/BPH report \\
\addlinespace[4pt]
Vaks et al., 2025, \emph{Journal of Radio Electronics} \cite{ref45_Vaks2025JRE}
& 12 PC without massage, 8 PC after massage, 10 BPH
& Group-level molecule lists and selected spectra; samples 13, 29 and 34 illustrated
& Full patient-level matrix not available in the material reviewed
& Extension with massage and BPH groups \\
\addlinespace[4pt]
Vaks et al., 2025, SFM lecture \cite{vaks2025studying}
& Patients with prostate tumors
& Pilot study, abstract and lecture details
& Preliminary pilot findings; full patient-level matrix not available in the material reviewed
& Conference presentation of pilot findings \\
\bottomrule
\end{tabularx}
\end{table}
\clearpage
 
\clearpage
\twocolumn
\bibliographystyle{unsrt}
\bibliography{ref}

@article{ref1_RecentGlobalPatterns,
  author = {Culp MB and others},
  title = {Recent global patterns in prostate cancer incidence and mortality rates},
  journal = {Eur Urol},
  year = {2020},
  volume = {77},
  number = {1},
  pages = {38-52},
  doi = {10.1016/j.eururo.2019.08.005},
  url = {https://doi.org/10.1016/j.eururo.2019.08.005},
}

@misc{ref2_IarcGlobalCancer,
  author = {{International Agency for Research on Cancer}},
  title = {Cancer Today: data visualization tools for exploring the global cancer burden},
  year = {2024},
  howpublished = {\url{https://gco.iarc.fr/today}},
  note = {Accessed 17 September 2026},
}

@book{ref3_RussianCancerReport,
  author = {Kaprin AD and Starinskiy VV and Shakhzadova AO and Lisichnikova IV},
  title = {Malignant tumors in Russia in 2022 (morbidity and mortality)},
  publisher = {MNIOI im. P.A. Gertsena -- filial FGBU NMITS radiologii Minzdrava Rossii},
  address = {Moscow},
  year = {2023},
  pages = {274},
  isbn = {978-5-85502-290-2},
}

@article{ref4_ProstateCancerUpdate,
  author = {Grubb RL},
  title = {Prostate Cancer: Update on Early Detection and New Biomarkers},
  journal = {Mo Med},
  year = {2018},
  volume = {115},
  number = {2},
  pages = {132-134},
  pmid = {30228704},
  pmcid = {PMC6139871},
  url = {https://pubmed.ncbi.nlm.nih.gov/30228704/},
}

@article{ref5_BiologyOfProstate,
  author = {Balk SP and Ko YJ and Bubley GJ},
  title = {Biology of prostate-specific antigen},
  journal = {J Clin Oncol},
  year = {2003},
  volume = {21},
  number = {2},
  pages = {383-391},
  doi = {10.1200/jco.2003.02.083},
  url = {https://doi.org/10.1200/jco.2003.02.083},
}

@article{ref6_DifferenDiagnosiOf,
  author = {Han C and Zhu L and Liu X and Ma S and Liu Y and Wang X},
  title = {Differential diagnosis of uncommon prostate diseases: combining mpMRI and clinical information},
  journal = {Insights Imaging},
  year = {2021},
  volume = {12},
  number = {1},
  pages = {79},
  doi = {10.1186/s13244-021-01024-3},
  url = {https://doi.org/10.1186/s13244-021-01024-3},
}

@article{ref7_DiagnostAbilityOf,
  author = {Wang W and Wang M and Wang L and others},
  title = {Diagnostic ability of \%p2PSA and prostate health index for aggressive prostate cancer: a meta-analysis},
  journal = {Sci Rep},
  year = {2014},
  volume = {4},
  pages = {5012},
  doi = {10.1038/srep05012},
  url = {https://doi.org/10.1038/srep05012},
}

@article{ref8_MoleculaPca3Diagnost,
  author = {Van Gils MP and Cornel EB and Hessels D and others},
  title = {Molecular PCA3 diagnostics on prostatic fluid},
  journal = {Prostate},
  year = {2007},
  volume = {67},
  number = {8},
  pages = {881-887},
  doi = {10.1002/pros.20564},
  url = {https://doi.org/10.1002/pros.20564},
}

@article{ref9_Pca3Tmprss2Erg,
  author = {Apolikhin OI and Sivkov AV and Efremov GD and others},
  title = {The first Russian experience of using PCA3 and TMPRSS2-ERG for prostate cancer diagnosis},
  journal = {Experimental and Clinical Urology},
  year = {2015},
  pages = {30-36},
  url = {https://ecuro.ru/en/article/first-russian-experience-using-pca3-and-tmprss2-erg-prostate-cancer-diagnosis},
}

@article{ref10_Mir888Is,
  author = {Lewis H and Lance R and Troyer D and others},
  title = {miR-888 is an expressed prostatic secretions-derived microRNA that promotes prostate cell growth and migration},
  journal = {Cell Cycle},
  year = {2014},
  volume = {13},
  number = {2},
  pages = {227-239},
  doi = {10.4161/cc.26984},
  url = {https://doi.org/10.4161/cc.26984},
}

@article{ref11_PotentiaUrinaryMirna,
  author = {Haj-Ahmad TA and Abdalla MA and Haj-Ahmad Y},
  title = {Potential urinary miRNA biomarker candidates for the accurate detection of prostate cancer among benign prostatic hyperplasia patients},
  journal = {J Cancer},
  year = {2014},
  volume = {5},
  number = {3},
  pages = {182-191},
  doi = {10.7150/jca.6799},
  url = {https://doi.org/10.7150/jca.6799},
}

@article{ref12_TheUtilityOf,
  author = {Stuopelyte K and Daniunaite K and Bakavicius A and others},
  title = {The utility of urine-circulating miRNAs for detection of prostate cancer},
  journal = {Br J Cancer},
  year = {2016},
  volume = {115},
  number = {6},
  pages = {707-715},
  doi = {10.1038/bjc.2016.233},
  url = {https://doi.org/10.1038/bjc.2016.233},
}

@article{ref13_DiagnostAndPrognost,
  author = {Fredsoe J and Rasmussen AKI and Thomsen AR and others},
  title = {Diagnostic and prognostic microRNA biomarkers for prostate cancer in cell-free urine},
  journal = {Eur Urol Focus},
  year = {2018},
  volume = {4},
  number = {6},
  pages = {825-833},
  doi = {10.1016/j.euf.2017.02.018},
  url = {https://doi.org/10.1016/j.euf.2017.02.018},
}

@article{ref14_AssessmeOfMir,
  author = {Dolotkazin DR and Averinskaya DA and Knyazev EN and others},
  title = {Assessment of miR-21-5p, miR-451a, and miR-144-3p level in urine in differential diagnosis of localized prostate cancer},
  journal = {Onkourologiya = Cancer Urology},
  year = {2024},
  volume = {20},
  number = {1},
  pages = {36-43},
  doi = {10.17650/1726-9776-2024-20-1-36-43},
  url = {https://doi.org/10.17650/1726-9776-2024-20-1-36-43},
}

@article{ref15_TheHumanUrine,
  author = {Bouatra S and Aziat F and Mandal R and Guo AC and Wilson MR and others},
  title = {The human urine metabolome},
  journal = {PLoS ONE},
  year = {2013},
  volume = {8},
  number = {9},
  pages = {e73076},
  doi = {10.1371/journal.pone.0073076},
  url = {https://doi.org/10.1371/journal.pone.0073076},
}

@article{ref16_ImpactOfExercise,
  author = {Deda O and Gika HG and others},
  title = {Impact of exercise and aging on rat urine and blood metabolome: an LC-MS based metabolomics longitudinal study},
  journal = {Metabolites},
  year = {2017},
  volume = {7},
  number = {1},
  pages = {10},
  doi = {10.3390/metabo7010010},
  url = {https://doi.org/10.3390/metabo7010010},
}

@article{ref17_SelectedIonFlow,
  author = {Smith D and Spanel P},
  title = {Selected ion flow tube mass spectrometry (SIFT-MS) for on-line trace gas analysis},
  journal = {Mass Spectrom Rev},
  year = {2005},
  volume = {24},
  pages = {661-700},
  doi = {10.1002/mas.20033},
  url = {https://doi.org/10.1002/mas.20033},
}

@article{ref19_TheUntargetUrine,
  author = {Llambrich M and Brezmes J and Cumeras R},
  title = {The untargeted urine volatilome for biomedical applications: methodology and volatilome database},
  journal = {Biol Proced Online},
  year = {2022},
  volume = {24},
  pages = {20},
  doi = {10.1186/s12575-022-00184-w},
  url = {https://doi.org/10.1186/s12575-022-00184-w},
}

@article{ref20_NovelMetaboliSignatur,
  author = {Yang B and Zhang C and Cheng S and Li G and Griebel J and Neuhaus J},
  title = {Novel metabolic signatures of prostate cancer revealed by 1H-NMR metabolomics of urine},
  journal = {Diagnostics},
  year = {2021},
  volume = {11},
  pages = {149},
  doi = {10.3390/diagnostics11020149},
  url = {https://doi.org/10.3390/diagnostics11020149},
}

@article{ref21_UntargetUrineMetaboli,
  author = {Ma Y and Zheng Z and Xu S and Attygalle A and Kim IY and Du H},
  title = {Untargeted urine metabolite profiling by mass spectrometry aided by multivariate statistical analysis to predict prostate cancer treatment outcome},
  journal = {Analyst},
  year = {2022},
  volume = {147},
  number = {13},
  pages = {3043-3054},
  doi = {10.1039/d2an00676f},
  url = {https://doi.org/10.1039/d2an00676f},
}

@article{ref22_The2017Terahert,
  author = {Dhillon SS and others},
  title = {The 2017 terahertz science and technology roadmap},
  journal = {J Phys D Appl Phys},
  year = {2017},
  volume = {50},
  pages = {043001},
  doi = {10.1088/1361-6463/50/4/043001},
  url = {https://doi.org/10.1088/1361-6463/50/4/043001},
}

@article{ref23_TeraMirRadiatio,
  author = {Pereira MF},
  title = {TERA-MIR radiation: materials, generation, detection and applications II},
  journal = {Opt Quantum Electron},
  year = {2015},
  volume = {47},
  pages = {815-820},
  doi = {10.1007/s11082-015-0146-x},
  url = {https://doi.org/10.1007/s11082-015-0146-x},
}

@article{10.1021/acsomega.3c10175,

    author = {Apostolakis, Apostolos and Aoust, Guillaume and Maisons, Gr{\'e}gory and Laurent, Ludovic and Pereira, Mauro Fernandes},

    title = {Photoacoustic Spectroscopy
Using a Quantum Cascade
Laser for Analysis of Ammonia in Water Solutions},

    journal = {ACS Omega},

    volume = {9},

    number = {17},

    pages = {19127-19135},

    year = {2024},

    month = {04},

    issn = {2470-1343},

    doi = {10.1021/acsomega.3c10175},

    url = {https://doi.org/10.1021/acsomega.3c10175},

    eprint = {https://pubs.acs.org/acsodf/article-pdf/9/17/19127/4797459/ao3c10175.pdf},

}

@Inbook{Cousin2022,
author="Cousin, Philippe
and Moumtzidou, Anastasia
and Karakostas, Anastasios
and Gounaridis, Lefteris
and Kouloumentas, Christos
and Pereira, Mauro Fernandes
and Apostolakis, Apostolos
and Gorrochategui, Paula
and Aoust, Guillaume
and Lebental, B{\'e}reng{\`e}re",
title="Improving Water Quality and Security with Advanced Sensors and Indirect Water Sensing Methods",
bookTitle="Instrumentation and Measurement Technologies for Water Cycle Management",
year="2022",
publisher="Springer International Publishing",
address="Cham",
pages="251--277",
isbn="978-3-031-08262-7",
doi="10.1007/978-3-031-08262-7_11",
url="https://doi.org/10.1007/978-3-031-08262-7_11"
}

@article{ref24_ApplicatOfMicrowav,
  author = {Vaks VL and Domracheva EG and Nikiforov SD and Sobakinskaya EA and Chernyaeva MB},
  title = {Application of microwave nonstationary spectroscopy for noninvasive medical diagnostics},
  journal = {Radiophys Quantum Electron},
  year = {2008},
  volume = {51},
  number = {6},
  pages = {493-498},
  doi = {10.1007/s11141-008-9049-z},
  url = {https://doi.org/10.1007/s11141-008-9049-z},
}

@article{ref25_SensingNitrilesWith,
  author = {Vaks V and Anfertev V and Chernyaeva M and Domracheva E and Yablokov A and Maslennikova A and others},
  title = {Sensing nitriles with {THz} spectroscopy of urine vapours from cancer patients subject to chemotherapy},
  journal = {Sci Rep},
  year = {2022},
  volume = {12},
  pages = {18117},
  doi = {10.1038/s41598-022-22783-z},
  url = {https://doi.org/10.1038/s41598-022-22783-z},
}

@article{ref26_ApplicatOfHigh,
  author = {Vaks V and Domracheva E and Chernyaeva M and Anfertev V and Maslennikova A and Atduev V and others},
  title = {Application of high-resolution terahertz gas spectroscopy for studying the composition of thermal decomposition products in human urine of prostate cancer patients},
  journal = {Appl Sci},
  year = {2024},
  volume = {14},
  pages = {1955},
  doi = {10.3390/app14051955},
  url = {https://doi.org/10.3390/app14051955},
}

@article{youden1950_index,
  author = {Youden WJ},
  title = {Index for rating diagnostic tests},
  journal = {Cancer},
  year = {1950},
  volume = {3},
  number = {1},
  pages = {32-35},
  doi = {10.1002/1097-0142(1950)3:1<32::AID-CNCR2820030106>3.0.CO;2-3},
  url = {https://doi.org/10.1002/1097-0142(1950)3:1<32::AID-CNCR2820030106>3.0.CO;2-3},
}

@article{ref27_HighResolutiTerahert,
  author = {Vaks VL and Anfertev VA and Balakirev VYu and Basov SA and Domracheva EG and Illyuk AV and others},
  title = {High resolution terahertz spectroscopy for analytical applications},
  journal = {Phys Usp},
  year = {2020},
  volume = {63},
  number = {7},
  pages = {708-720},
  doi = {10.3367/ufne.2019.07.038613},
  url = {https://doi.org/10.3367/ufne.2019.07.038613},
}

@article{ref28_OnThePossibil,
  author = {Vaks VL and Anfertev VA and Chernyaeva MB and Domracheva EG and Pripolzin SI and Baranov AN and others},
  title = {On the possibility of advancement of the non-stationary gas spectroscopy method realized by using fast frequency sweep mode up the terahertz frequency range},
  journal = {Radiophys Quantum Electron},
  year = {2023},
  volume = {65},
  number = {10},
  pages = {760-774},
  doi = {10.1007/s11141-023-10255-x},
  url = {https://doi.org/10.1007/s11141-023-10255-x},
}

@article{ref29_ApplicatOfA,
  author = {Vaks VL and Domracheva EG and Chernyaeva MB and Anfertev VA and Maslennikova AV and Zheleznyak AV and others},
  title = {Application of a high-resolution terahertz gas spectroscopy method to compositional analysis of thermal decomposition products of human fluids (urine)},
  journal = {J Opt Technol},
  year = {2022},
  volume = {89},
  number = {4},
  pages = {243-249},
  doi = {10.1364/jot.89.000243},
  url = {https://doi.org/10.1364/jot.89.000243},
}

@article{ref30_GiantControllGigahert,
  author = {Pereira MF and Anfertev V and Shevchenko Y and others},
  title = {Giant controllable gigahertz to terahertz nonlinearities in superlattices},
  journal = {Sci Rep},
  year = {2020},
  volume = {10},
  pages = {15950},
  doi = {10.1038/s41598-020-72746-5},
  url = {https://doi.org/10.1038/s41598-020-72746-5},
}

@article{ref31_AnalyticExpressiFor,
  author = {Pereira MF},
  title = {Analytical expressions for numerical characterization of semiconductors per comparison with luminescence},
  journal = {Materials},
  year = {2017},
  volume = {11},
  number = {1},
  pages = {2},
  doi = {10.3390/ma11010002},
  url = {https://doi.org/10.3390/ma11010002},
}

@article{ref32_HarmonicGeneratiIn,
  author = {Pereira MF},
  title = {Harmonic generation in biased semiconductor superlattices},
  journal = {Nanomaterials},
  year = {2022},
  volume = {12},
  pages = {1504},
  doi = {10.3390/nano12091504},
  url = {https://doi.org/10.3390/nano12091504},
}

@article{ref33_TerahertGeneratiBy,
  author = {Pereira MF and Anfertev and Zubelli JP and Vaks VL},
  title = {Terahertz generation by gigahertz multiplication in superlattices},
  journal = {J Nanophotonics},
  year = {2017},
  volume = {11},
  number = {4},
  pages = {046022},
  doi = {10.1117/1.jnp.11.046022},
  url = {https://doi.org/10.1117/1.jnp.11.046022},
}

@article{ref34_ProgressInAnalytic,
  author = {Al-Ateqi A and Pereira MF},
  title = {Progress in analytical solutions for high order harmonic generation in semiconductor superlattice multipliers},
  journal = {Opt Quantum Electron},
  year = {2023},
  volume = {55},
  pages = {1287},
  doi = {10.1007/s11082-023-05555-5},
  url = {https://doi.org/10.1007/s11082-023-05555-5},
}

@article{ref35_JplSpectralCatalog,
  author = {Pickett HM and Poynter RL and Cohen EA and Delitsky ML and Pearson JC and Muller HSP},
  title = {Submillimeter, Millimeter, and Microwave Spectral Line Catalog},
  journal = {Journal of Quantitative Spectroscopy and Radiative Transfer},
  year = {1998},
  volume = {60},
  number = {5},
  pages = {883-890},
  doi = {10.1016/S0022-4073(98)00091-0},
  url = {https://spec.jpl.nasa.gov/},
}

@article{ref37_UrinaryMetaboliBiomarke,
  author = {Abina A and Korosec T and Puc U and Jazbinsek M and Zidansek A},
  title = {Urinary metabolic biomarker profiling for cancer diagnosis by terahertz spectroscopy: review and perspective},
  journal = {Photonics},
  year = {2023},
  volume = {10},
  pages = {1051},
  doi = {10.3390/photonics10091051},
  url = {https://doi.org/10.3390/photonics10091051},
}

@article{ref38_GlycolalAReactive,
  author = {Nagai R and Matsumoto K and Ling X and Suzuki H and Araki T and Horiuchi S},
  title = {Glycolaldehyde, a reactive intermediate for advanced glycation end products, plays an important role in the generation of an active ligand for the macrophage scavenger receptor},
  journal = {Diabetes},
  year = {2000},
  volume = {49},
  number = {10},
  pages = {1714-1723},
  doi = {10.2337/diabetes.49.10.1714},
  url = {https://doi.org/10.2337/diabetes.49.10.1714},
}

@article{lin2021_quantSchema,
  author = {Lin W and Conway LP and Vujasinovic M and Lohr JM and Globisch D},
  title = {Chemoselective and Highly Sensitive Quantification of Gut Microbiome and Human Metabolites},
  journal = {Angewandte Chemie International Edition},
  year = {2021},
  volume = {60},
  pages = {23232-23240},
  doi = {10.1002/anie.202107101},
  url = {https://doi.org/10.1002/anie.202107101},
}

@article{ref39_TheWarburgEffect,
  author = {Banerjee A and others},
  title = {The Warburg effect in cancer: therapeutic implications and early detection},
  journal = {Int J Cancer Res Ther},
  year = {2025},
  volume = {10},
  number = {1},
  pages = {1-4},
  doi = {10.33140/ijcrt.10.01.05},
  url = {https://doi.org/10.33140/ijcrt.10.01.05},
}

@article{ref40_SerumMalondiaMda,
  author = {Lepara Z and others},
  title = {Serum malondialdehyde (MDA) levels as a potential biomarker of cancer progression in patients with bladder cancer},
  journal = {Rom J Intern Med},
  year = {2020},
  volume = {58},
  number = {3},
  pages = {146-152},
  doi = {10.2478/rjim-2020-0008},
  url = {https://doi.org/10.2478/rjim-2020-0008},
}

@article{ref53_UrinaryMalondiaMda,
  author = {Toto A and Wild P and Graille M and Turcu V and Creze C and Hemmendinger M and Sauvain JJ and Bergamaschi E and Guseva Canu I and Hopf NB},
  title = {Urinary malondialdehyde (MDA) concentrations in the general population: a systematic literature review and meta-analysis},
  journal = {Toxics},
  year = {2022},
  volume = {10},
  number = {4},
  pages = {160},
  doi = {10.3390/toxics10040160},
  url = {https://doi.org/10.3390/toxics10040160},
}

@article{ref41_PhenylalanineBenzaldehyde,
  author = {Nierop Groot MN and de Bont JAM},
  title = {Conversion of phenylalanine to benzaldehyde initiated by an aminotransferase in Lactobacillus plantarum},
  journal = {Applied and Environmental Microbiology},
  year = {1998},
  volume = {64},
  number = {8},
  pages = {3009-3013},
  doi = {10.1128/AEM.64.8.3009-3013.1998},
  url = {https://doi.org/10.1128/AEM.64.8.3009-3013.1998},
}

@article{ref42_IdentifiOfA,
  author = {Lima AR and Pinto J and Azevedo AI and others},
  title = {Identification of a biomarker panel for improvement of prostate cancer diagnosis by volatile metabolic profiling of urine},
  journal = {Br J Cancer},
  year = {2019},
  volume = {121},
  pages = {857-868},
  doi = {10.1038/s41416-019-0585-4},
  url = {https://doi.org/10.1038/s41416-019-0585-4},
}

@article{ref43_ProstateCancerScreenin,
  author = {Deev V and others},
  title = {Prostate cancer screening using chemometric processing of GC-MS profiles obtained in the headspace above urine samples},
  journal = {J Chromatogr B Analyt Technol Biomed Life Sci},
  year = {2020},
  volume = {1155},
  pages = {122298},
  doi = {10.1016/j.jchromb.2020.122298},
  url = {https://doi.org/10.1016/j.jchromb.2020.122298},
}

@article{ref44_MechanisAndConseque,
  author = {Delanghe S and Delanghe JR and Speeckaert R and Van Biesen W and Speeckaert MM},
  title = {Mechanisms and consequences of carbamoylation},
  journal = {Nat Rev Nephrol},
  year = {2017},
  volume = {13},
  pages = {580-593},
  doi = {10.1038/nrneph.2017.103},
  url = {https://doi.org/10.1038/nrneph.2017.103},
}

@article{ref46_NonInvasiveUrinary,
  author = {Perez-Rambla C and Puchades-Carrasco L and Garcia-Flores M and Rubio-Briones J and Lopez-Guerrero JA and Pineda-Lucena A and others},
  title = {Non-invasive urinary metabolomic profiling discriminates prostate cancer from benign prostatic hyperplasia},
  journal = {Metabolomics},
  year = {2017},
  volume = {13},
  pages = {52},
  doi = {10.1007/s11306-017-1194-y},
  url = {https://doi.org/10.1007/s11306-017-1194-y},
}

@article{ref47_1hNmrBased,
  author = {Zniber M and Lamminen T and Taimen P and Bostrom PJ and Huynh TP},
  title = {1H-NMR-based urine metabolomics of prostate cancer and benign prostatic hyperplasia},
  journal = {Heliyon},
  year = {2024},
  volume = {10},
  number = {7},
  pages = {e28949},
  doi = {10.1016/j.heliyon.2024.e28949},
  url = {https://doi.org/10.1016/j.heliyon.2024.e28949},
}

@article{ref48_UrinaryVolatileOrganic,
  author = {Khalid T and Aggio R and White P and De Lacy Costello B and Persad R and Al-Kateb H and others},
  title = {Urinary volatile organic compounds for the detection of prostate cancer},
  journal = {PLoS ONE},
  year = {2015},
  volume = {10},
  number = {11},
  pages = {e0143283},
  doi = {10.1371/journal.pone.0143283},
  url = {https://doi.org/10.1371/journal.pone.0143283},
}

@article{ref49_MajorOdorantsReleased,
  author = {Kim KH and Jahan SA and Kabir E},
  title = {Major odorants released as urinary volatiles by urinary incontinent patients},
  journal = {Sensors},
  year = {2013},
  volume = {13},
  number = {7},
  pages = {8523-8533},
  doi = {10.3390/s130708523},
  url = {https://doi.org/10.3390/s130708523},
}

@article{ref50_DeterminOfFormalde,
  author = {Takeuchi A and Takigawa T and Abe M and Kawai T and Endo Y and Yasugi T and Endo G and Ogino K},
  title = {Determination of formaldehyde in urine by headspace gas chromatography},
  journal = {Bull Environ Contam Toxicol},
  year = {2007},
  volume = {79},
  pages = {1-4},
  doi = {10.1007/s00128-007-9172-0},
  url = {https://doi.org/10.1007/s00128-007-9172-0},
}

@article{ref52_UrinaryMethanolAnd,
  author = {Berode M and Sethre T and Laubli T and Savolainen H},
  title = {Urinary methanol and formic acid as indicators of occupational exposure to methyl formate},
  journal = {Int Arch Occup Environ Health},
  year = {2000},
  volume = {73},
  pages = {410-414},
  doi = {10.1007/s004200000160},
  url = {https://doi.org/10.1007/s004200000160},
}

@article{ref54_UrinaryInflammaAnd,
  author = {Jiang YH and Lee J and Kuo HC and Wu YH},
  title = {Urinary inflammatory and oxidative stress biomarkers as indicators for the clinical management of benign prostatic hyperplasia},
  journal = {Int J Mol Sci},
  year = {2025},
  volume = {26},
  number = {13},
  pages = {6516},
  doi = {10.3390/ijms26136516},
  url = {https://doi.org/10.3390/ijms26136516},
}

@INPROCEEDINGS{ref44_Vaks2024IEEE,
  title           = "Analysis of thermal decomposition products of urine at
                     prostate cancer and benign prostatic hyperplasia by
                     terahertz high-resolution gas spectroscopy",
  booktitle       = "2024 {IEEE} 9th {All-Russian} Microwave Conference ({RMC})",
  author          = "Vaks, Vladimir and Chernyaeva, Mariya and Atduev, Vagif
                     and Anfertev, Vladimir and Maslennikova, Anna and
                     Rodionov, Mikhail and Domracheva, Elena and Atduev, Kurban",
  publisher       = "IEEE",
  pages           = "228--232",
  month           =  nov,
  year            =  2024,
  conference      = "2024 IEEE 9th All-Russian Microwave Conference (RMC)",
  location        = "Moscow, Russian Federation"
}

@article{ref45_Vaks2025JRE,
  author = {Vaks, V. L. and Anfertev, V. A. and Domracheva, E. G. and
            Chernyaeva, M. B. and Maslennikova, A. V. and Atduev, V. A. and
            Rodionov, M. A. and Atduev, K. A.},
  title = {Analysis of content of thermal decomposition products of urine at
           prostate cancer and benign prostatic hyperplasia by terahertz
           nonstationary gas spectroscopy},
  journal = {Journal of Radio Electronics},
  year = {2025},
  number = {1},
  doi = {10.30898/1684-1719.2025.1.5},
  note = {In Russian}
}

@inproceedings{vaks2025studying,
  author    = {Vaks, V. L. and Atduev, V. A. and Maslennikova, A. V. and Domracheva, E. G. and Chernyaeva, M. B. and Anfertev, V. A. and Atduev, K. A. and Rodionov, M. A. and Pereira, M. F.},
  title     = {Studying the composition of urine thermal decomposition products of patients with prostate tumors (pilot study) by high-resolution terahertz gas spectroscopy},
  booktitle = {Saratov Fall Meeting 2025},
  year      = {2025},
  month     = {September 29--October 3},
  note      = {Invited lecture},
  address   = {Saratov, Russia}
}

@article{Tang2025UrineMetabolomics,
  author  = {Tang, Chenye and Guo, Xiao and Zou, Qian and
             Wang, Xinghao and Sheng, Jian and Shen, Dan and
             Sun, Chun and Li, Shuo and Shen, Ruilin},
  title   = {Untargeted metabolomics revealed urinary metabolic pattern
             for discriminating prostate cancer from benign prostatic
             hyperplasia in Chinese participants},
  journal = {Frontiers in Oncology},
  year    = {2025},
  volume  = {15},
  pages   = {1604169},
  doi     = {10.3389/fonc.2025.1604169}
}

@article{PanachNavarrete2026UrineNMR,
  author  = {Panach-Navarrete, Jorge and
             Gonz{\'a}lez-Marrachelli, Vannina and
             Morales-Tatay, Jos{\'e} Manuel and
             Garc{\'i}a-Morata, Francisco and
             Sales-Maicas, Mar{\'i}a {\'A}ngeles and
             Monle{\'o}n-Salvado, Daniel and
             Mart{\'i}nez-Jabaloyas, Jos{\'e} Mar{\'i}a},
  title   = {Urine metabolic analysis as a noninvasive method to
             diagnose prostate cancer},
  journal = {Urologic Oncology: Seminars and Original Investigations},
  year    = {2026},
  volume  = {44},
  number  = {2},
  pages   = {125.e1--125.e10},
  doi     = {10.1016/j.urolonc.2025.10.015}
}

@article{Mahmud2026PaperSpray,
  author  = {Mahmud, Iqbal and Pinto, Frederico G. and
             Garrett, Timothy J.},
  title   = {Rapid Noninvasive Classification of Prostatic Disease
             Using Paper Spray Ionization Mass Spectrometry
             ({PSI-MS})-Based Nontargeted Metabolomics},
  journal = {Analytical Chemistry},
  year    = {2026},
  volume  = {98},
  number  = {18},
  pages   = {13203--13210},
  doi     = {10.1021/acs.analchem.6c00795}
}

@article{Yoo2026OlfactoryNanodiscs,
  author  = {Yoo, Jin and Kim, Yerin and Kang, Ku and
             Ahn, Young Jin and Bang, Sungun and Koo, Kyo Chul and
             Park, Tai Hyun},
  title   = {Urine-Based Noninvasive Detection of Prostate Cancer
             Using Human Olfactory Receptor-Embedded Nanodiscs},
  journal = {ACS Sensors},
  year    = {2026},
  volume  = {11},
  number  = {5},
  pages   = {4121--4131},
  doi     = {10.1021/acssensors.6c00794}
}

@article{Wang2026SerumSHAP,
  author  = {Wang, Zijie and Xin, Jialu and He, Qiuyan and
             Xu, Shutong and Wu, Jinghan and Yang, Fang and
             Dong, Liang},
  title   = {Serum Untargeted Metabolomics Integrated with
             {SHAP}-Based Machine Learning for Multiclass Stratification
             of Prostate Cancer, Prostatitis, and Benign Prostatic
             Hyperplasia},
  journal = {Metabolites},
  year    = {2026},
  volume  = {16},
  number  = {4},
  pages   = {237},
  doi     = {10.3390/metabo16040237}
}

@article{ApostolakisPereira+2020+3941+3952,
url = {https://doi.org/10.1515/nanoph-2020-0155},
title = {Superlattice nonlinearities for Gigahertz-Terahertz generation in harmonic multipliers},
title = {},
author = {Apostolos Apostolakis and Mauro F. Pereira},
pages = {3941--3952},
volume = {9},
number = {12},
journal = {Nanophotonics},
doi = {doi:10.1515/nanoph-2020-0155},
year = {2020},
lastchecked = {2026-10-06}
}

@article{Oriaku:17,
author = {C. I. Oriaku and M. F. Pereira},
journal = {J. Opt. Soc. Am. B},
number = {2},
pages = {321--328},
publisher = {Optica Publishing Group},
title = {Analytical solutions for semiconductor luminescence including Coulomb correlations with applications to dilute bismides},
volume = {34},
month = {Feb},
year = {2017},
url = {https://opg.optica.org/josab/abstract.cfm?URI=josab-34-2-321},
doi = {10.1364/JOSAB.34.000321},
}

@article{10.1117/1.JNP.13.036017,
author = {Apostolos Apostolakis and Mauro F. Pereira},
title = {{Potential and limits of superlattice multipliers coupled to different input power sources}},
volume = {13},
journal = {Journal of Nanophotonics},
number = {3},
publisher = {SPIE},
pages = {036017},
year = {2019},
doi = {10.1117/1.JNP.13.036017},
URL = {https://doi.org/10.1117/1.JNP.13.036017}
}

@article{10.1063/1.5050917,
    author = {Apostolakis, A. and Pereira, Mauro F.},
    title = {Controlling the harmonic conversion efficiency in semiconductor superlattices by interface roughness design},
    journal = {AIP Advances},
    volume = {9},
    number = {1},
    pages = {015022},
    year = {2019},
    month = {01},
    issn = {2158-3226},
    doi = {10.1063/1.5050917},
    url = {https://doi.org/10.1063/1.5050917},
    eprint = {https://pubs.aip.org/aip/adv/article-pdf/doi/10.1063/1.5050917/12899940/015022_1_online.pdf},
}

\end{document}